\documentclass{IEEE-Con-Sys-mag}

\jvol{XX}
\jnum{XX}
\paper{8}
\jmonth{June}
\jname{IEEE CONTROL SYSTEMS}
\pubyear{2020}

\usepackage{tikz}
\usetikzlibrary{arrows.meta,calc}
\definecolor{lightblue}{HTML}{33CCCC}
\newcommand{\legendbox}[1]{\node[anchor=north east,draw=black!60,fill=white,fill opacity=0.9,text opacity=1,inner sep=2.5pt,font=\scriptsize,align=left] at #1}
\usepackage[]{hyperref} 
\usepackage{balance}

\def\ds{\displaystyle}
\newcommand{\rev}[1]{#1}
\newcommand{\newmat}[1]{#1}
\newcommand{\revon}{}
\definecolor{BLUE}{rgb}{0,0,1}

\begin{document}

\title{A systems theoretic look at the Riemann zeta function\stitle{}}

\author{{A}lessandro Astolfi}

\affil{Computer, Electrical and Mathematical Science and Engineering Division, 
King Abdullah University of Science and Technology (KAUST), 
Thuwal, 23955, Saudi Arabia, alessandro.astolfi@kaust.edu.sa}

\maketitle

\bgroup\def\thefootnote{}\def\footnotemark{}\footnotetext{\fontsize{6.5pt}{9pt}\selectfont\hspace*{-8pt}\textit{This work has been submitted to the IEEE for possible publication. Copyright may be transferred without notice, after which this version may no longer be accessible.}}\egroup

\begin{pullquote}
Systems theory and number theory are two very different disciplines: we illustrate how the former can contribute to the latter.
\end{pullquote}

\chapterinitial{S}ystems theory and number theory are two very different disciplines. The former deals with the study of control systems, 
by providing powerful tools for their analysis and design, the latter is one of the richest areas of mathematics, 
devoted to the study of natural numbers and their building blocks, the prime numbers. A pivotal role in (analytic) number theory, 
see “Number theory” for more details on the various branches of number theory,
is played by the Riemann zeta function, see \cite{Riemann1859},

\begin{equation}\label{zeta_f}
\zeta(s) = \sum_{n=1}^{\infty} \frac{1}{n^s} = \frac{1}{1^s} + \frac{1}{2^s} + \frac{1}{3^s} + \cdots , \qquad \Re(s) >1.
\end{equation}
Some key features of the Riemann zeta function are encoded in its poles and zeros. It is well-known that the Riemann zeta function
has a single pole, at $s=1$, with residue equal to $1$, and countably many zeros: 
the {\em trivial\ \/}zeros at $-2$, $-4$, $-6$, $\cdots$; and the {\em non-trivial\ \/}zeros located in the so-called critical strip 
$0 < \Re(s) <1$. The location of the non-trivial zeros forms the subject of the celebrated Riemann Hypothesis, 
one of the most important unsolved problems in mathematics, 
which conjectures that all non-trivial zeros of the Riemann zeta function have real part equal to $\frac{1}{2}$, 
that is, they belong to what is called the {\em critical line}.

% ---------- Sidebar Number theory -----------

\begin{sidebar}{Number theory}\label{sidebar:numb_th}
\sdbarinitial{N}umber theory, see, for example, \cite{HardyWright2008, IwaniecKowalski2004anth, montgomery},
is one of the oldest and most influential branches of mathematics, 
devoted to the study of integers and their arithmetic properties. 
Classical topics include divisibility, prime numbers, congruences, Diophantine equations, and arithmetic functions. 
Over time, the subject has developed into several major areas, including elementary, analytic, algebraic, geometric, and computational number theory. 
Each approach employs distinct techniques and viewpoints, together providing a comprehensive understanding of arithmetic phenomena. 
\begin{itemize} 
\item \textbf{Elementary Number Theory:} Studies the arithmetic properties of integers using elementary methods. 
        Typical topics include divisibility, prime numbers, congruences, greatest common divisors, and Diophantine equations. 
\item \textbf{Analytic Number Theory:} Applies tools from real and complex analysis to investigate arithmetic questions. 
         It is particularly concerned with the distribution of prime numbers, arithmetic functions, and the properties of the Riemann zeta function 
         and Dirichlet $L$-functions. 
\item \textbf{Algebraic Number Theory:} Extends the theory of integers to algebraic number fields and their rings of integers. Important concepts include ideals, units, class groups, and factorization in algebraic structures. 
\item \textbf{Geometric Number Theory:} Employs geometric methods to solve arithmetic problems. The geometry of numbers studies lattices, convex bodies, and Diophantine approximation, providing geometric insights into number-theoretic questions. 
\item \textbf{Computational Number Theory:} Focuses on the design and analysis of efficient algorithms for arithmetic computations. It includes primality testing, integer factorization, modular arithmetic, and elliptic curve algorithms, with significant applications in cryptography and computer security. 
\end{itemize} 
\end{sidebar}
% ------

The aim of this paper is to show that,
by interpreting the Riemann zeta function, and its (multiplicative) inverse, as the transfer functions of linear, 
infinite dimensional systems in open-loop and closed-loop architectures, elementary systems theory 
(plus standard results in complex analysis and calculus) allows deriving important properties of the Riemann zeta function
and of the Mertens function, which is widely studied in number theory. 
This perspective also yields algorithms to compute the Riemann zeta function
using simple operations, which do not involve the use of special functions, with explicit error bounds. 
These algorithms coincide with the classical (first-order) Euler--Maclaurin summation formula, 
see \cite[Chapter 6]{Edwards1974}, which is thus given a systems theoretic derivation. 
Finally, the error bounds allow determining, numerically, regions of the critical strip which do not contain zeros.

To achieve these objectives simple facts are exploited. 
The main one is that the region of convergence of the Laplace transform of the step response of a system 
bounds the location of its dominant pole: if the step response grows at most as $e^{\sigma_0 t}$ (possibly multiplied by a polynomial in $t$), 
then the transfer function is analytic in the half-plane $\Re(s)>\sigma_0$, and the degree of the polynomial, plus one, bounds the 
multiplicity of the poles with real part equal to $\sigma_0$.
Note that, with the change of variable $x=e^t$, the Laplace transform becomes the Mellin transform, 
which is a standard tool in analytic number theory \cite{Titchmarsh1986}: 
the systems theoretic perspective provides therefore a dictionary between the two disciplines, 
which yields elementary and transparent derivations of several classical results. 
In terms of the Riemann zeta function, only its Dirichlet series, the Dirichlet series of its inverse, 
the fact that it is meromorphic, and, to exploit the symmetry of its zeros, the functional equation are used.

\rev{No new result in number theory is established: 
the facts about the Riemann zeta function and the Mertens function which are derived in what follows are classical, 
and the algorithms coincide with known summation formulas. The contribution is the systems theoretic viewpoint itself, 
namely the dictionary between transfer functions, step responses, and regions of convergence on one side, 
and Dirichlet series, summatory functions, and abscissae of convergence on the other, 
which makes these facts, their proofs, and the explicit error bounds accessible with the tools of an undergraduate 
course in linear systems, and which provides the logical structure (exclusion of cells, a robust counting argument, 
and the symmetry of the zeros) of a certified algorithm for the isolation of the zeros, 
described in the Appendix. The paper is therefore expository in nature.}

% ---------------
\begin{summary}
\summaryinitial{E}lementary systems theory 
(plus standard results in complex analysis and calculus) allows deriving important properties of the Riemann zeta function
and of some key functions which are widely studied in number theory. 
\end{summary}
% --------------

% ---------- Sidebar Zeta -----------
\begin{sidebar}{The Riemann zeta function}\label{sidebar:Riemann}
\sdbarinitial{T}he Riemann zeta function, introduced by Euler as a function of a real variable,
see \cite{Riemann1859,Titchmarsh1986,Edwards1974,Hardy1914,Conrey2003},
is defined as
$$
\zeta(s) = \sum_{n=1}^{\infty} \frac{1}{n^s}, \qquad \Re(s) >1.
$$ 
The function $\zeta(s)$ can be analytically continued to a meromorphic function on the entire complex plane. 
Its only singularity is a simple pole at $s=1$ with residue equal to $1$. The function has an essential singularity at infinity.

The Riemann zeta function satisfies the functional equation
\begin{equation}\label{eq:functional_eq}
\zeta(s) = 2^s \pi^{s-1} \sin\!\left(\frac{\pi s}{2}\right) \Gamma(1-s) \zeta(1-s),
\end{equation}
where $\Gamma$ denotes the $\Gamma$ function \cite{artin1964gamma,olver2010nist}.
The functional equation relates the values of $\zeta(s)$ and $\zeta(1-s)$ 
and reveals a remarkable symmetry about the critical line $\Re(s)=\frac{1}{2}$.

The Riemann zeta function is intimately related to the prime numbers via the Euler product formula, 
valid for $\Re(s) >1,$
$$
\zeta(s) = \prod_{p \ \text{prime}} \frac{1}{1-p^{-s}}.
$$
This formula establishes a fundamental connection between the Riemann zeta function and the distribution of primes.

The zeros of the Riemann zeta function play a central role in analytic number theory. 
The function has trivial zeros at the negative even integers
(these can be seen directly from the functional equation), that is at 
$$
s=-2,-4,-6,\ldots,
$$
while all other zeros lie in the critical strip
$$
0<\Re(s)<1.
$$
The famous Riemann Hypothesis asserts that every nontrivial zero has real part equal to $\frac{1}{2}$.

The Riemann zeta function appears throughout mathematics, including analytic number theory, spectral theory, 
probability, mathematical physics, and the study of special functions. 
\end{sidebar}
% ------

% ---------- Sidebar Gamma -----------
\begin{sidebar}{The Gamma function}\label{sidebar:Gamma}
\sdbarinitial{T}he Gamma function \cite{artin1964gamma} is one of the most important special functions in mathematics 
and serves as a continuous extension of the factorial function to complex numbers. 
For $\Re(s)>0$, it is defined by the Euler integral 
$$
\Gamma(s)=\int_{0}^{\infty} t^{\,s-1}e^{-t}\,dt. 
$$ 
It satisfies the fundamental functional equation 
$\Gamma(s+1)=s\,\Gamma(s),$ 
which implies that $\Gamma(n+1)=n!$ for every nonnegative integer $n$. 
The Gamma function plays a central role in analysis, probability theory, differential equations, and number theory, 
where it appears in the study of the Riemann zeta function, Dirichlet $L$-functions, and various integral transforms.
\end{sidebar}
% ------

\section{Notation and definitions}

To provide concise statements we introduce some (standard) notation and definitions.
$\mathbb{N}=\{1,2,3,\ldots\}$ denotes the set of natural numbers;
$\mathbb{R}$ denotes the set of real numbers;
$\mathbb{R}_{\ge 0}$ denotes the set of non-negative real numbers;
$\mathbb{C}$ denotes the set of complex numbers;
$\Re(s)$ and $\Im(s)$ denote the real and imaginary parts of $s\in \mathbb{C}$, respectively.

Given a function $f: \mathbb{R}_{\ge 0} \to \mathbb{R}$, 
its (unilateral) Laplace transform is denoted as ${\cal L}f=F(s)$, whereas
${\cal L}^{-1} F(s)$ denotes the inverse Laplace transform of $F(s)$, see \cite{doetsch1974laplace,widder1941laplace}.
Impulses at $t=0$ are included in the domain of integration (that is, the lower limit of integration is $0^-$).
$\delta$ denotes the Dirac $\delta$ function. Recall that ${\cal L}\delta = 1$.
$H$ denotes the Heaviside function. Recall that ${\cal L}H = \ds\frac{1}{s}$. Finally, recall that 
${\cal L}\left[f(t-\tau)H(t-\tau)\right] = e^{-\tau s} {\cal L} f$, for all $\tau \ge 0$.
The notation $\log$ denotes the natural log function, 
the notation $\lfloor x \rfloor$ denotes the floor function, that is the largest integer not larger than $x$, 
and $\{x\} = x - \lfloor x \rfloor$ denotes the fractional part of $x$. 
Additional special functions, which often occur in (analytic) number theory are defined in the Sidebars.
\rev{Finally, a function $W: \mathbb{C} \to \mathbb{C}$ is meromorphic if it is analytic on the whole complex plane 
with the exception of a set of isolated points, the poles, at each of which it is of the form $W(s) = \frac{h(s)}{(s-p)^m}$, 
for some integer $m \ge 1$ and some function $h$ which is analytic and non-zero at $p$. 
A meromorphic function is thus the natural generalization of a rational function: 
its zeros and poles are isolated and have finite multiplicities, but they may be countably many, 
as is the case for the transfer functions of delay systems. The Riemann zeta function and its multiplicative inverse are meromorphic.}

\begin{definition}\label{def:dominant}
Let $W: \mathbb{C} \to \mathbb{C}$ be a meromorphic function, not identically zero. 
Let ${\cal Z} \subset \mathbb{C}$ and ${\cal P} \subset \mathbb{C}$ be the sets of the (finite) zeros and poles, respectively, of $W$. 
Let 
$$
\alpha = \sup_{z \in {\cal Z}} {\Re} (z), \qquad \beta = \sup_{p \in {\cal P}} {\Re} (p),
$$
with the convention that the supremum of the empty set is $-\infty$.
$\alpha$ is called the dominant zero of $W$ and $\beta$ is called the dominant pole of $W$.
In addition, if $W$ has zeros (respectively, poles) with real part equal to $\alpha$ (respectively, $\beta$) 
and the multiplicities of these zeros (respectively, poles) are bounded, 
the largest of these multiplicities is called the multiplicity of the dominant zero (respectively, pole).
\end{definition}

\rev{The terms dominant zero and dominant pole are borrowed from systems theory, in which the dominant poles of a transfer function 
are the right-most ones: Definition~\ref{def:dominant} 
retains the term, but refers to the abscissa, that is to the real part, of the right-most zeros and poles, 
which is the quantity relevant for the analysis in what follows (in analytic number theory 
this is the abscissa of convergence, or of absolute convergence, of the associated Dirichlet series).}
The numbers $\alpha$ and $\beta$ need not be the real part of a zero or of a pole, respectively, 
of the function $W$, since the suprema in Definition~\ref{def:dominant} need not be attained.
For example, for the Riemann zeta function it is known that $\alpha \in [1/2,1]$, since there are zeros on the critical line \cite{Hardy1914} 
and there are no zeros with real part larger than or equal to one \cite{Titchmarsh1986}, 
but it is not known whether the supremum is attained, unless the Riemann Hypothesis holds.

\section{The open loop zeta system}

% ---------------
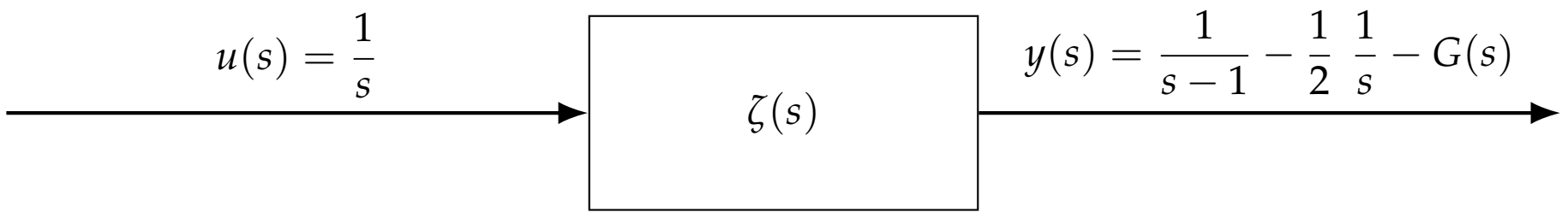
\begin{figure*}[htb]
\begin{center}
\begin{tikzpicture}[
    block/.style={draw,rectangle,minimum width=3cm,minimum height=1.5cm,align=center},
    >={Latex}
    ]
\node[block] (B) {$\zeta(s)$}; \draw[->, thick] (-6,0) -- (B.west)node[midway, above] {$\ds u(s) =  \frac{1}{s}$};
\draw[->, thick] (B.east) -- (6,0) node[midway, above] {$\ds y(s) = \frac{1}{s-1} - \frac{1}{2}\ \frac{1}{s} - G(s)$};
\end{tikzpicture}
\caption{The open loop zeta system.}
\label{fig:Zeta}
\end{center}
\end{figure*}
% ---------------

\begin{figure}[h]
\centerline{\includegraphics[width=18.0pc]{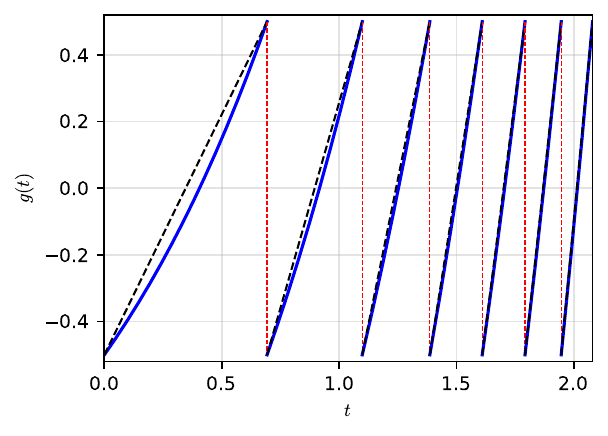}}
\caption{The graph of the function $g$. Note that in each interval $[\log n, \log(n+1))$ the function $g$ is convex and its graph
             is below that of the straight dashed black lines from the points $(\log n, -1/2)$ to the points $(\log(n+1), 1/2)$, 
             for all $n \in \mathbb{N}$. 
             The dashed red lines highlight the discontinuity of $g$.}
              \label{fig:plot_gt} 
\end{figure}

Let the Riemann zeta function be the transfer function of a linear (infinite dimensional) system with input $u$ and output 
$y$ (see Figure \ref{fig:Zeta}) and note that the Riemann zeta function has an explicit inverse Laplace transform, that is
$$
{\cal L}^{-1} \zeta(s) = \sum_{n \in \mathbb{N}} \delta(t-\log n),
$$
for all $t \in \mathbb{R}_{\ge 0}$.
This allows interpreting the Riemann zeta function as the transfer function of a system composed of
 the parallel interconnection of countable delay systems,  with delay $\tau_n = \log n$, with $n \in \mathbb{N}$. 
The output response of the system in Figure \ref{fig:Zeta} with input the Heaviside function,
that is the step response of the zeta system, can be explicitly computed 
to give, for all $t \in \mathbb{R}_{\ge 0}$,
$$
y(t) = \lfloor e^t \rfloor = e^t  - \frac{1}{2} - 
\bigg(\overbrace{
                  e^t -  \lfloor e^t \rfloor - \frac{1}{2}}^{g(t)}
\bigg) = e^t - \frac{1}{2} - g(t).
$$
The function 
$$
g(t) = e^t -  \lfloor e^t \rfloor - \frac{1}{2} = \left\{e^t\right\} - \frac{1}{2},
$$
see Figure \ref{fig:plot_gt}, 
is piecewise continuous, right continuous, and such that 
$-1/2 \le g(t) < 1/2$, for all $t \in \mathbb{R}_{\ge 0}$, hence $g$ is
Laplace transformable, with Laplace transform $G(s)$ with a well-defined region of convergence.

\subsection{Region of convergence of $G(s)$}

To compute the region of convergence of the Laplace transform of $g$ note, to begin with, that the inequality
$$
|g(t)| \le \frac{1}{2},
$$
for all $t \in \mathbb{R}_{\ge 0}$, implies that the integral
$$
\ds \int_0^\infty e^{-s t} g(t) dt 
$$
converges absolutely for all $s$ such that $\Re(s)>0$. 
It is shown below that, because of the oscillatory nature of $g$, the integral converges, although not absolutely, 
in the larger half-plane $\Re(s)>-1$. To this end, consider first the value at $s=0$, and let
$$
G(0) = \ds  \int_0^\infty g(t)  dt 
        = \sum_{n \in \mathbb{N}} \int_{\log n}^{\log (n+1)} \left(e^t - n -\frac{1}{2}\right) dt 
                                      = \sum_{n \in \mathbb{N}}  I_n
$$
and note that 
$$
I_n = 1-\left(n+\frac12\right) \log\!\left(1+\frac1n\right),
$$
and that $I_n<0$, for all $n \in \mathbb{N}$.  
A simple way to see that $I_n<0$, for all $n \in \mathbb{N}$ is to note, from the graphs in Figure~\ref{fig:plot_gt}, 
that the straight dashed-black lines are described by the equation
$$
l(t)= -\frac{1}{2} + \frac{t-\log n}{\log\!\left(\frac{n+1}{n}\right)}, 
$$
for $t \in[ \log n, \log(n+1)]$, for all $n \in \mathbb{N}$,
and that 
$$
\int_{\log n}^{\log(n+1)} l(t)\,dt = 0, \quad n\in\mathbb{N}.
$$
Hence, since $g$ is always (strictly) below the dashed line, its integral in the same intervals is negative.

Consider now the expansion, valid for all $n \ge 1$ (for $n=1$ the series converges conditionally),
$$
\log\!\left(1+\frac1n\right) = \frac{1}{n} - \frac{1}{2 n^2} + \frac{1}{3 n^3}  - \cdots
$$
which yields 
$$
\begin{array}{ccl}
I_n & = & \ds \sum_{m=2}^{\infty} (-1)^{m+1} \frac{m-1}{2m(m+1)} \,n^{-m}  \\[4mm]
      & = & \ds -\frac{1}{12n^2} + \frac{1}{12n^3} -\frac{3}{40n^4} +\cdots . \\[2mm]
\end{array}      
$$
This implies that the series $\ds \sum_{n \in \mathbb{N}}  I_n$ is absolutely convergent. 
Moreover, for all $T \in [\log n, \log(n+1)]$,
$$
\left| \int_{\log n}^{T} g(t)\,dt \right| \le \frac{1}{2} \log\!\left(1+\frac{1}{n}\right),
$$
which tends to zero as $n$ tends to $+\infty$, hence the improper integral defining $G(0)$ converges and its value coincides with the sum of the series.
To compute this sum note that, for all $N \ge 2$,
$$
\begin{array}{ccl}
\ds \sum_{n=1}^{N-1}  I_n & = & \ds N-1 - \sum_{n=1}^{N-1} \left(n+\frac12\right) \Big( \log(n+1) - \log n \Big) \\[6mm]
                         & = & \ds N-1 - \left(N+\frac12\right) \log N + \log N!,
\end{array}
$$
and that, by Stirling's formula, $\log N! = N \log N - N + \frac{1}{2} \log(2 \pi N) + o(1)$. As a result,
$$
G(0) =  \sum_{n \in \mathbb{N}}  I_n = \frac{1}{2} \log(2 \pi) - 1 \approx -0.08106146679.
$$
Consider now
$$
G(s) = \int_0^\infty e^{-s t} g(t) dt,
$$
that is the Laplace transform of $g$. Using the same technique exploited for evaluating $G(0)$ 
yields, for all $s$ such that $\Re(s)>0$,
$$
G(s) = \sum_{n \in \mathbb{N}}  J_n(s),
$$
with 
$$
J_n(s) = \ds \frac{\ds \left(n+\frac{1+s}{2}\right)(n+1)^{-s} - \left(n+\frac{1-s}{2}\right)n^{-s}} {s(1-s)}.
$$
Note, now, that an expansion in powers of $1/n$ yields
$$
J_n(s)  = -\frac{1+s}{12}\, n^{-s-2} + O\!\left((1+|s|)^2\, n^{-\Re(s)-3}\right),
$$
uniformly for $s$ in compact sets (for $s=0$ this is consistent with the expansion of $I_n=J_n(0)$). 
Hence the series $\sum_{n \in \mathbb{N}} J_n(s)$ converges absolutely and uniformly on compact subsets of the half-plane $\Re(s)>-1$, 
and defines an analytic function in this half-plane. In addition, for $\Re(s) = \sigma \in (-1,0]$ and $T \in [\log n, \log(n+1)]$,
$$
\left| \int_{\log n}^{T} e^{-st} g(t)\,dt \right| \le \frac{1}{2}\, (n+1)^{-\sigma} \log\!\left(1+\frac{1}{n}\right),
$$
which tends to zero as $n$ tends to $+\infty$. Therefore, the improper integral defining $G(s)$ converges, and coincides with 
$\sum_{n \in \mathbb{N}} J_n(s)$, for all $s$ such that $\Re(s)>-1$.
In summary, $G(s)$ is analytic in the half-plane $\Re(s)>-1$, and the integral defining it converges absolutely for $\Re(s)>0$.

\subsection{An alternative description of the Riemann zeta function}

For $\Re(s)>1$ the step response of the system in Figure \ref{fig:Zeta}, that is $y(t) = e^t - \frac{1}{2} - g(t)$, 
is Laplace transformable, and its Laplace transform is
$$
\frac{\zeta(s)}{s} = \frac{1}{s-1}  - \frac{1}{2}\ \frac{1}{s} - G(s).
$$
This yields the alternative description of the Riemann zeta function
\begin{equation}\label{eq:zeta_s}
\zeta(s) = \frac{s}{s-1} - \frac{1}{2} -  s G(s),
\end{equation}
which holds, to begin with, for $\Re(s)>1$. 
Since the right-hand side of equation (\ref{eq:zeta_s}) is analytic in the half-plane $\Re(s)>-1$, with the exception of $s=1$, 
by uniqueness of the analytic continuation, equation (\ref{eq:zeta_s}) holds for all $s$ such that $\Re(s) > -1$ and $s \neq 1$. 
With the change of variable $x=e^t$, equation (\ref{eq:zeta_s}) becomes
$$
\zeta(s) = \frac{s}{s-1} - \frac{1}{2} -  s \int_1^\infty \left( \{x\} - \frac{1}{2} \right) x^{-s-1} \, dx,
$$
which is a classical representation of the Riemann zeta function, obtained by Euler--Maclaurin summation, 
see \cite[Section 2.1]{Titchmarsh1986}.
Since $G$ is analytic at $s=0$, the term $s G(s)$ vanishes at $s=0$.
This representation of the Riemann zeta function allows concluding the following facts:

\medskip

\begin{enumerate}
\item $\beta=1$ is the dominant pole of the Riemann zeta function, with multiplicity $1$ and residue equal to $1$;
\item $\zeta(0) =  -1/2$;
\item $\zeta^\prime(0) = -1 - G(0)  = -\ds \frac{1}{2} \log(2 \pi)$.
\end{enumerate}

\medskip

The first fact follows from $\frac{s}{s-1} = 1 + \frac{1}{s-1}$ and from analyticity of $G$ in the half-plane $\Re(s)>-1$, 
whereas the second and third facts follow by evaluating equation (\ref{eq:zeta_s}) and its derivative at $s=0$.
These are, of course, classical facts. Their derivation, however, only requires the elementary computations above, 
whereas several textbook derivations of $\zeta(0)$ and $\zeta^\prime(0)$ rely on the functional equation and on the
properties of the Gamma function.

\subsection{Computation of the Riemann zeta function}

The description of the Riemann zeta function in equation (\ref{eq:zeta_s}) allows computing the function for all $s \in \mathbb{C}$, with
$\Re(s) > -1$, and $s \neq 1$, by numerical evaluation of  $G(s)$. 
Observe that while the evaluation of the Riemann zeta function for $\Re(s) > 1$ can be undertaken by using the very definition 
of the function, its evaluation for $\Re(s) \le 1$, and in particular in the critical strip $0<\Re(s) < 1$, 
requires an analytic continuation, typically obtained by means of Euler--Maclaurin summation, of the functional equation, 
or of the Riemann--Siegel formula \cite{Edwards1974}. 
\rev{It is demonstrated, in what follows, that the systems theoretic perspective yields, with elementary arguments, 
an algorithm with an explicit error bound. The algorithm coincides with first-order Euler--Maclaurin summation, 
hence it is neither new nor, at large heights, competitive with the Riemann--Siegel formula, which requires $O(\sqrt{|\Im(s)|})$ 
operations per evaluation against $O(|\Im(s)|^{4/3})$, see the discussion following inequality (\ref{eq:boundR}). 
The novelty is in the derivation, which uses only the bound $|g(t)| \le 1/2$ and an integration by parts; 
in a two-sided error bound which is tight (see Figure \ref{fig:alpha_Rn_alpha}); and in the fact that the bound holds 
in the whole half-plane $\Re(s) > -1$, which is what is needed for the exclusion test of 
inequality (\ref{eq:test}) and for the algorithm in the Appendix.}

\subsubsection{The interval $(0,1)$}

\begin{figure}[htb]
\centerline{\includegraphics[width=18.0pc]{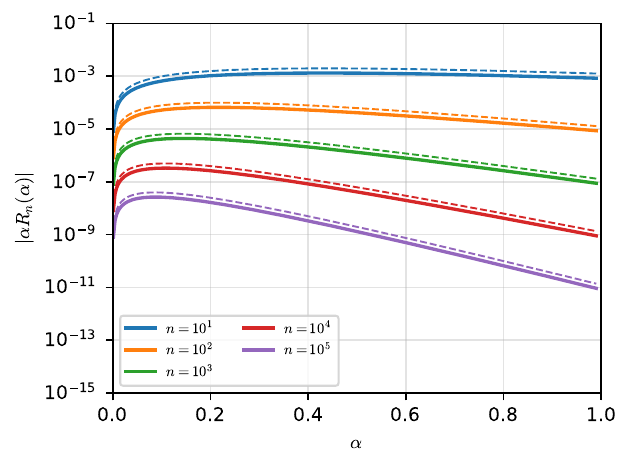}}
\caption{The modulus of the approximation error $\alpha R_n(\alpha)$ (solid lines) and its bound $\frac{\alpha}{8} n^{-\alpha-1}$ 
             (dashed lines) as a function of $\alpha \in (0,1)$, for $n=10, 10^2, 10^3, 10^4, 10^5$.
             Note the logarithmic scale on the $y$-axis.}
              \label{fig:alpha_Rn_alpha} 
\end{figure}

To begin with, consider the evaluation of $G(s)$ on the real segment $(0,1)$, 
that is the function $G(\alpha)$ for $\alpha \in (0,1)$, defined as
$$
G(\alpha) = \int_0^\infty e^{-\alpha t} g(t) dt. 
$$
The value $\alpha=0$ is excluded since it has already been shown 
that $G(0) = \frac{1}{2} \log(2 \pi) -1$.
Note that, for any $n \in \mathbb{N}$,
\begin{equation}\label{eq:G_alpha_approx}
G(\alpha) = \overbrace{\int_0^{\log n} e^{-\alpha t} g(t) dt}^{G_n(\alpha)} \ + \ 
\overbrace{\int_{\log n}^\infty e^{-\alpha t} g(t) dt}^{R_n(\alpha)},
\end{equation}
where $G_n(\alpha)$ yields the numerical approximation of $G(\alpha)$ and $R_n(\alpha)$ 
describes the approximation error.
The function $G_n(\alpha)$ can be computed  iteratively as 
\begin{equation}\label{eq:recursion}
\begin{array}{ccl}
G_{k+1}(\alpha)  & = & \ds G_k(\alpha)  \ + \ 
                                     \ds \frac{(k+1)^{1-\alpha}-k^{1-\alpha}}{1-\alpha} \ -\\[4mm]
                           &    &\ds \frac{k+\frac12}{\alpha} \Big(k^{-\alpha}-(k+1)^{-\alpha}\Big),\\[2mm]
\end{array} 
\end{equation}
with 
$$
G_{1}(\alpha) = \int_0^{\log 1} e^{-\alpha t} g(t) dt = 0,
$$ 
The approximation error $R_n(\alpha)$ admits the simple bound 
$|R_n(\alpha)| \le \frac{1}{2} \frac{n^{-\alpha}}{\alpha}$, obtained by using the bound $|g(t)| \le 1/2$, for all $t \in \mathbb{R}_{\ge 0}$.
A much sharper bound is obtained by exploiting the fact that the signal $e^t g(t)$ has a bounded integral.
To see this, let
$$
p(t) = \int_0^t e^\tau g(\tau)\, d\tau,
$$
and note that, for $\tau \in [\log k, \log(k+1))$, $e^\tau g(\tau)\,d\tau = (x-k-\frac{1}{2})\,dx$, with $x=e^\tau$. Hence
$$
p(t) = \frac{1}{2} \left( \left\{e^t\right\}^2 - \left\{e^t\right\} \right),
$$
which implies that $p(\log k) = 0$, for all $k \in \mathbb{N}$, and that $-1/8 \le p(t) \le 0$, for all $t \in \mathbb{R}_{\ge 0}$.
Since $g(t) = e^{-t} \dot p(t)$ (for all $t$ which are not the logarithm of an integer), integrating by parts yields
\begin{equation}\label{eq:R_ibp}
R_n(\alpha) = \int_{\log n}^\infty e^{-(\alpha+1)t} \dot p(t)\, dt = (\alpha+1) \int_{\log n}^\infty e^{-(\alpha+1)t} p(t)\, dt,
\end{equation}
where the boundary terms vanish because $p(\log n) = 0$ and $p$ is bounded. As a result,
$$
-\frac{1}{8}\, n^{-\alpha-1} \le R_n(\alpha) \le 0.
$$
Using equation (\ref{eq:G_alpha_approx}) and the alternative description in equation (\ref{eq:zeta_s}) yields
\begin{equation}\label{eq:zeta_alpha_approx}
\zeta(\alpha) = \overbrace{\frac{\alpha}{\alpha-1} -\frac{1}{2} - \alpha G_n(\alpha)}^{\hat\zeta_n(\alpha)} - 
                \overbrace{\alpha R_n(\alpha)}^{{\cal E}_n(\alpha)},
\end{equation}
where $\hat\zeta_n(\alpha)$ describes the approximation of $\zeta(\alpha)$ and ${\cal E}_n(\alpha)$ is the approximation error, 
which is such that
$$
\hat\zeta_n(\alpha) \le \zeta(\alpha) \le \hat\zeta_n(\alpha) + \frac{\alpha}{8}\, n^{-\alpha-1},
$$
that is, $\hat\zeta_n(\alpha)$ is a lower bound for $\zeta(\alpha)$.
A direct computation, using the recursion in equation (\ref{eq:recursion}), shows that
\begin{equation}\label{eq:zeta_hat_closed}
\hat\zeta_n(\alpha) = \sum_{k=1}^{n-1} k^{-\alpha} + \frac{n^{1-\alpha}}{\alpha-1} + \frac{n^{-\alpha}}{2},
\end{equation}
that is, $\hat\zeta_n(\alpha)$ is the first-order Euler--Maclaurin approximation of $\zeta(\alpha)$, see \cite[Chapter 6]{Edwards1974}, and that
$$
{\cal E}_n(\alpha) = \alpha R_n(\alpha) = \frac{n^{1-\alpha}}{\alpha-1} + \frac{n^{-\alpha}}{2} - \zeta(\alpha,n),
$$
where $\zeta(\alpha,n)$ denotes the Hurwitz zeta function, which is defined as $\sum_{m=n}^{\infty} m^{-\alpha}$ for $\alpha>1$ 
and by analytic continuation for $\alpha<1$.
A graph of the modulus of the approximation error ${\cal E}_n(\alpha)$, together with its bound, is displayed in 
Figure \ref{fig:alpha_Rn_alpha} as a function of $\alpha \in (0,1)$ and for increasing values of $n$. 
Note that the bound is tight and that a {\em modest\ \/}value of $n$ yields a very good numerical approximation: 
since $\alpha n^{-\alpha} \le 1/(e \log n)$, the selection $n=10^4$ guarantees $|{\cal E}_n(\alpha)| < 5 \cdot 10^{-7}$, 
for all $\alpha \in (0,1)$.

\begin{figure}[htb]
\centerline{\includegraphics[width=18.0pc]{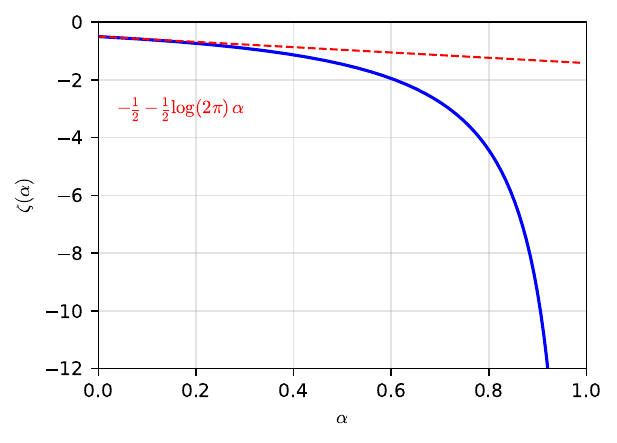}}
\caption{The Riemann zeta function along the real axis in the interval $(0,1)$, computed as $\hat\zeta_n(\alpha)$ with $n=1000$. 
             The dashed line is the tangent to $\zeta(\alpha)$ at $\alpha = 0$.}
              \label{fig:zeta_real} 
\end{figure}

The value of the Riemann zeta function on the real axis for $\alpha \in (0,1)$ is displayed in Figure \ref{fig:zeta_real}. 
Observe the pole at $\alpha = 1$ and the value $\zeta(0) = -1/2$.
The straight line is the tangent at $\alpha=0$, obtained using the facts established in the previous section.

Finally, note that $G(\alpha)$ is well defined for all $\alpha \in (0,1]$ and that 
the special case $\alpha=1$ yields the simplified iteration
$$
\ds G_{n+1}(1) = G_n(1) + \log\!\left(1+\frac{1}{n}\right) - \frac{n+\frac{1}{2}}{n(n+1)},
$$
with $G_1(1)=0$, and the bound $-\frac{1}{8n^2} \le R_n(1) \le 0$. 
Moreover, comparing the expansion $\zeta(s) = \frac{1}{s-1} + \gamma + O(s-1)$, where $\gamma$ is the Euler--Mascheroni constant, 
with equation (\ref{eq:zeta_s}) yields $G(1) = \frac{1}{2} - \gamma$. Hence the above iteration provides the two-sided bound
$$
\frac{1}{2} - G_n(1) < \gamma \le \frac{1}{2} - G_n(1) + \frac{1}{8n^2},
$$
\rev{which yields an approximation of $\gamma$ to any prescribed accuracy, with an explicit bound on the error. 
Note that $\frac{1}{2} - G_n(1) = \sum_{k=1}^{n-1} \frac{1}{k} + \frac{1}{2n} - \log n$, that is, the iteration 
is the classical definition of $\gamma$ as the limit of $\sum_{k=1}^{n} \frac{1}{k} - \log n$, 
accelerated by the term $\frac{1}{2n}$: the error decreases as $n^{-2}$ instead of $n^{-1}$. 
This is not competitive with the dedicated algorithms used to compute $\gamma$ to billions of digits, see \cite{Brent1980}, 
and is mentioned only as an illustration of the two-sided bound.}

\subsubsection{The critical strip}

The accurate evaluation of the Riemann zeta function, with an explicit error bound, 
in the critical strip is key to explore, numerically, the Riemann Hypothesis. In this respect we recall 
that the alternative description of the Riemann zeta function given in equation (\ref{eq:zeta_s}) 
is valid in the critical strip and requires only the numerical evaluation of the function $G(s)$ 
to yield the approximation
\begin{equation}\label{eq:zeta_approx_s}
\hat \zeta_n(s) = \frac{s}{s-1} -\frac{1}{2} - s \ G_n(s),
\end{equation}
with 
$$
G_n(s) = \sum_{k=1}^{n-1} J_k(s).
$$
The function  $G_n(s)$ can be computed iteratively using the recursion (which is the same as the one in 
equation (\ref{eq:recursion}) with $s$ instead of $\alpha$)
$$
\begin{array}{ccl}
G_{k+1}(s) & = & \ds G_k(s) + \frac{(k+1)^{1-s}-k^{1-s}}{1-s} \ - \\[4mm]
                  &     & \ds  \left(k+\frac12\right)\frac{k^{-s}-(k+1)^{-s}}{s}, 
\end{array}                  
$$
with $G_1(s) = 0$, or, equivalently, $\hat\zeta_n(s)$ is given by equation (\ref{eq:zeta_hat_closed}) with $s$ in place of $\alpha$.
The approximation error is $\zeta(s) - \hat\zeta_n(s) = - s R_n(s)$ and, repeating the integration by parts in 
equation (\ref{eq:R_ibp}) with $s$ in place of $\alpha$, yields, for all $s$ such that $\Re(s) > -1$, 
\begin{equation}\label{eq:boundR}
|s R_n(s)| \le \frac{|s|\, |s+1|}{8 \left(\Re(s)+1\right)}\, n^{-\Re(s)-1}.
\end{equation}
For complex $s$ the bound obtained using only $|g(t)| \le 1/2$ is $|sR_n(s)| \le \frac{|s|}{2\Re(s)}\, n^{-\Re(s)}$, and 
therefore deteriorates linearly with $|\Im(s)|$. The bound in inequality (\ref{eq:boundR}) is much sharper, 
provided $n$ is large compared with $|\Im(s)|$, which is consistent with the classical behavior of Euler--Maclaurin summation \cite{Edwards1974}.

% ---------------
\begin{figure*}[htb]
\begin{center}
\centerline{\begin{tikzpicture}
\node[anchor=south west,inner sep=0] (img) at (0,0) {\includegraphics[width=40.0pc]{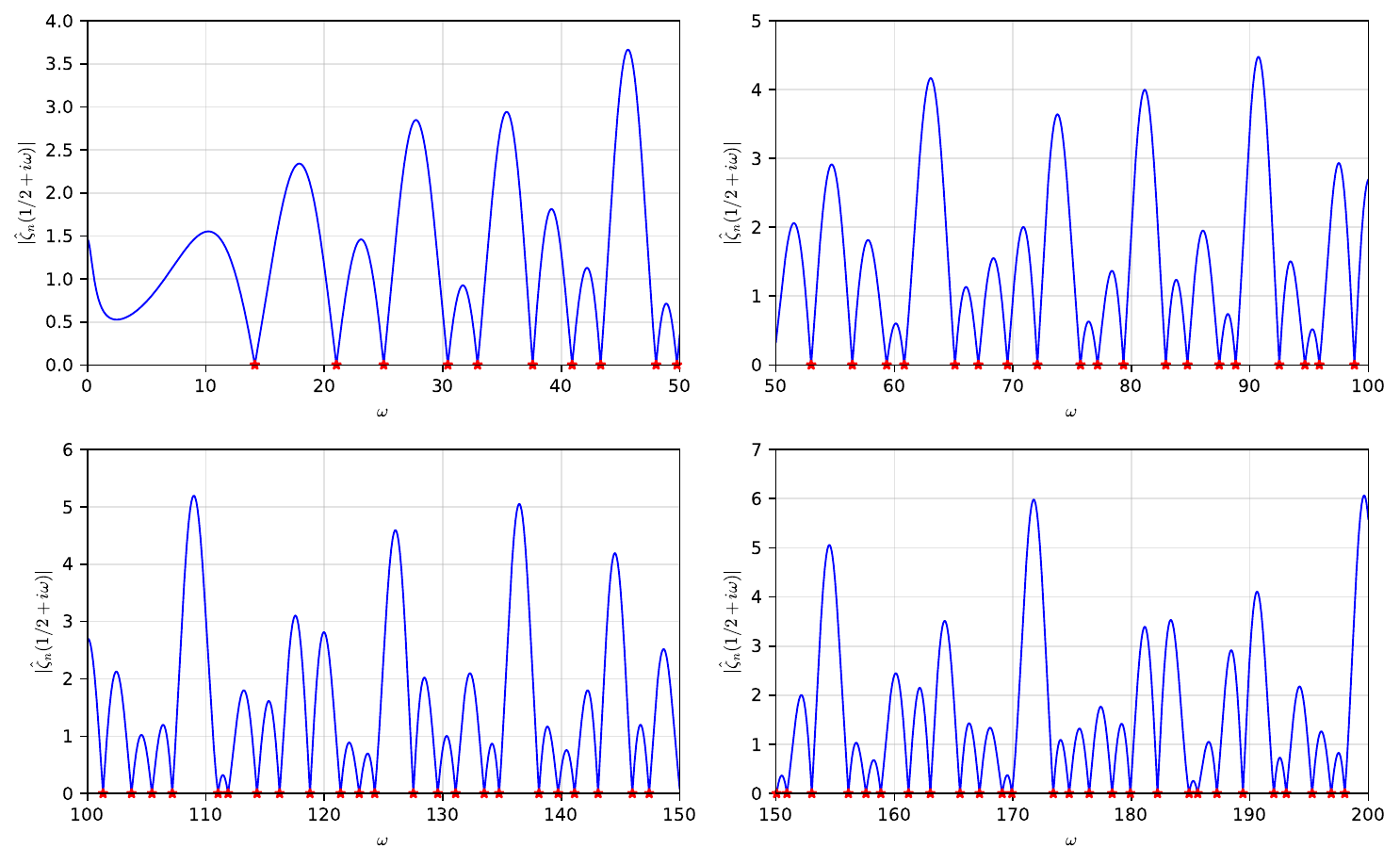}};
\legendbox{($(img.north west)+(8.2,-0.45)$)} {{\color{blue}\rule[2pt]{10pt}{0.8pt}}\ $|\hat\zeta_n(1/2+i\omega)|$\\ {\color{red}$\star$}\ zeros of $\zeta$};
\end{tikzpicture}}
\caption{The modulus of the Riemann zeta function on the critical line, evaluated numerically as $|\hat\zeta_n(1/2 + i\omega)|$, with $n=10^4$. 
             The red markers indicate the locations of the zeros of the Riemann zeta function.
\label{fig:critical_line}}
\end{center}
\end{figure*}
% ---------------

% ---------------
\begin{figure*}[htb]
\begin{center}
\centerline{
\includegraphics[width=22.0pc]{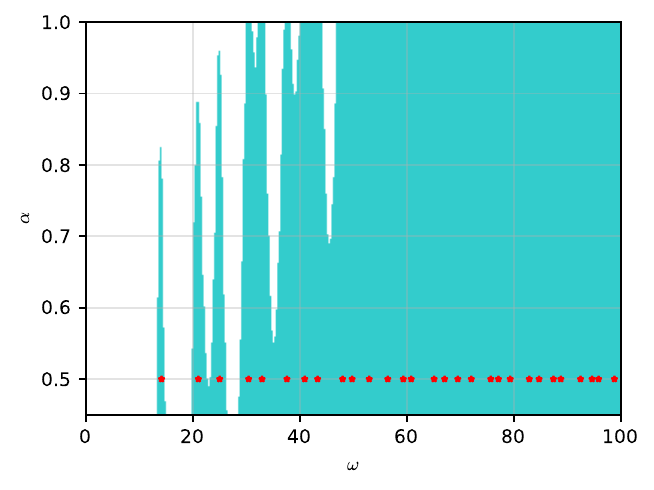} 
\begin{tikzpicture}
\node[anchor=south west,inner sep=0] (img) at (0,0) {\includegraphics[width=22.0pc]{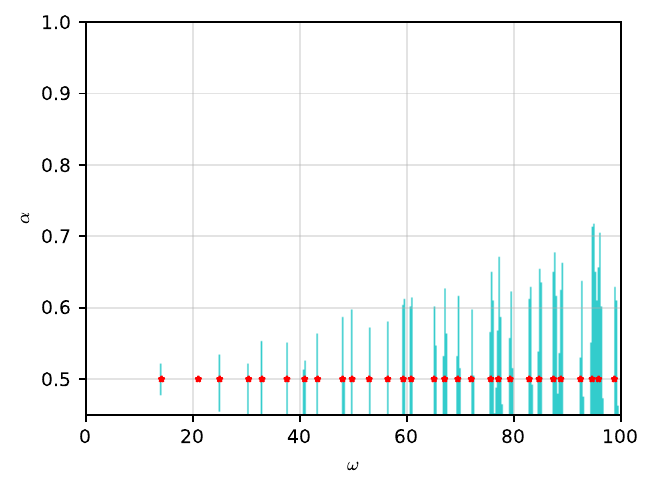}};
\legendbox{($(img.north east)+(-0.5,-0.4)$)} {{\color{lightblue}\rule[-1pt]{10pt}{6pt}}\ not excluded by inequality (\ref{eq:test})\\ {\color{red}$\star$}\ zeros of $\zeta$};
\end{tikzpicture}
}

\vspace{2mm}
\centerline{
\includegraphics[width=22.0pc]{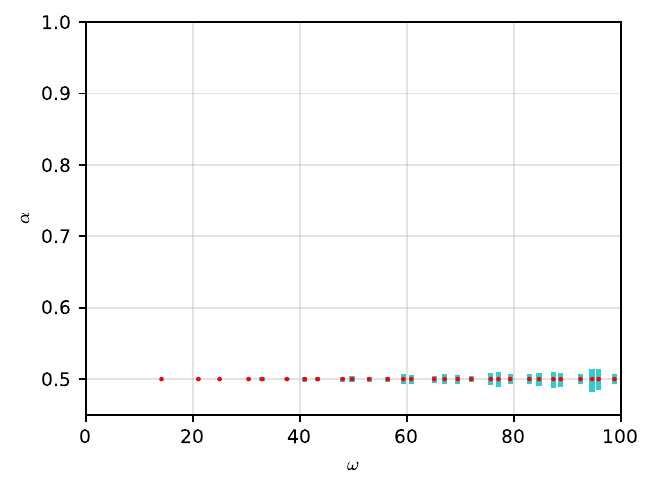}
\includegraphics[width=22.0pc]{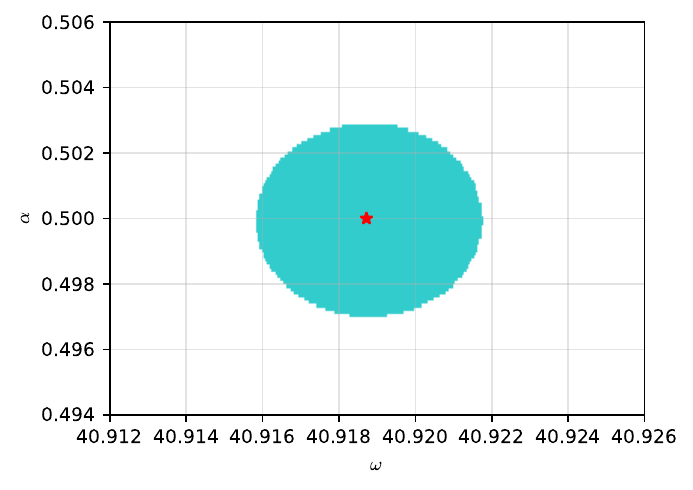}
}
\caption{Regions which may contain nontrivial zeros (light blue), that is points which are not excluded by the test in inequality (\ref{eq:test}), 
for $\alpha \in [0.45, 1]$ and $\omega \in [0,100]$, 
for $n=10$ (top-left); $n=100$ (top-right); $n=1000$ (bottom-left); 
and zoomed plot, for $n=1000$, around the zero with $\omega \approx 40.9187$ (bottom-right), 
which highlights the shape of the regions which may contain the nontrivial zeros. 
The red markers indicate the locations of the zeros of the Riemann zeta function.
\label{fig:nontrivial_zeros}}
\end{center}
\end{figure*}
% ---------------

To study the zeros of the Riemann zeta function in the critical strip note that 
the functional equation (\ref{eq:functional_eq}) implies that the nontrivial zeros are symmetric with respect to the 
line $\Re(s) = \frac{1}{2}$ (this does not imply that the Riemann zeta function has the same symmetry), and that, 
since $\zeta(\bar s) = \overline{\zeta(s)}$, the zeros are also symmetric with respect to the real axis. 
We therefore use  the approximation in equation (\ref{eq:zeta_approx_s}) 
for $\Re(s) \in \left[\frac{1}{2} - \epsilon,1 \right]$, with $\epsilon >0$ and small, and $\Im(s) \ge 0$, to approximate the location
of the zeros of the Riemann zeta function and to determine regions which do not contain zeros.

\begin{figure*}[htb]
\begin{center}
\centerline{\includegraphics[width=36.0pc,trim=61 23 56 34,clip]{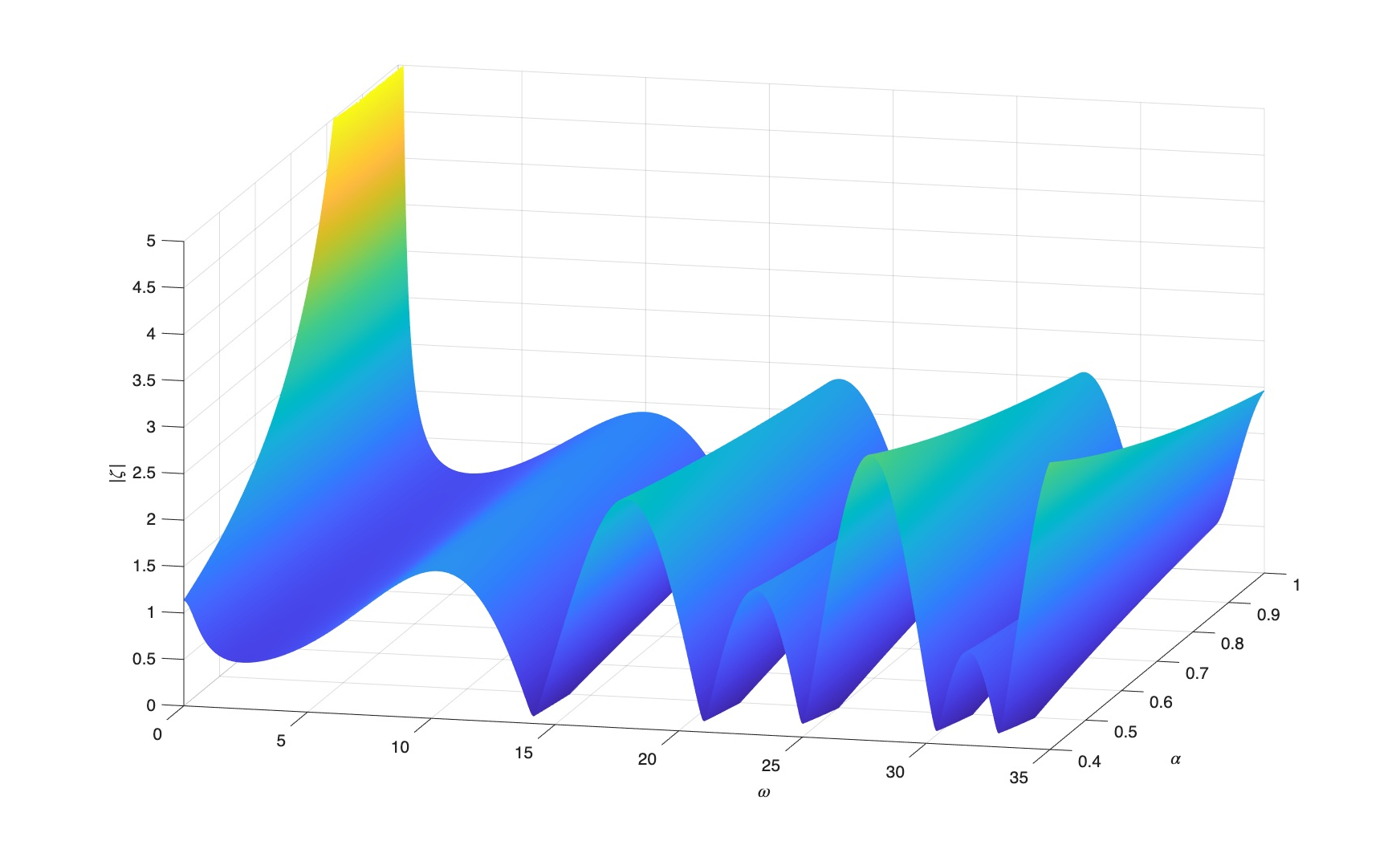}}
\caption{The modulus of the Riemann zeta function, evaluated numerically as $|\hat\zeta_n(\alpha + i\omega)|$, with $n=1000$, 
for $\alpha \in [0.4,1]$ and $\omega \in [0,35]$. Values larger than $5$, which occur near the pole at $s=1$, are not displayed.
\label{fig:modulus_zeta}}
\end{center}
\end{figure*}

To begin with, consider the critical line, defined as the line $s = \frac{1}{2} + i \omega$, for $\omega\in \mathbb{R}$. 
(Note that the classical literature on the Riemann zeta function uses the notation 
$s = \frac{1}{2} + i t$, for $t \in \mathbb{R}$. 
The variable $\omega$, instead of $t$, is used herein since this is standard notation in systems theory and also 
because $t$ is used to denote time in the paper.)
On the critical line, the bound in inequality (\ref{eq:boundR}) becomes $\frac{1}{12} |s|\, |s+1|\, n^{-3/2}$, which, 
for $n=10^4$ and $\omega \in [0,200]$, is smaller than $3.4 \cdot 10^{-3}$. Hence a graph of the modulus of
$\hat\zeta_n\left(\frac{1}{2} + i \omega\right)$ as a function of $\omega$, for $n=10^4$, provides an accurate 
insight on the location of zeros on the critical line. 
Figure \ref{fig:critical_line}, where the $\star$ markers indicate the location of the zeros of the Riemann zeta function and
the range $\omega \in [0,200]$ contains the first 79 nontrivial zeros of the Riemann zeta function, 
demonstrates the validity of the proposed approximation approach: 
all zeros on the critical line up to $\omega=200$ are identified by the approximation.
A complementary view is given in Figure \ref{fig:modulus_zeta}, which displays $|\hat\zeta_n(\alpha + i\omega)|$ 
for $\alpha \in [0.4,1]$ and $\omega \in [0,35]$: the first five nontrivial zeros appear as the points in which the surface 
touches zero, all on the line $\alpha = 1/2$, whereas the large values near $\omega = 0$ and $\alpha=1$ are due to the pole at $s=1$.

% ---------------
% ---------------

Note that, on its own, the proposed approximation cannot be used to prove that a point in the complex plane is a zero 
of the Riemann zeta function, or that a zero lies on the critical line: this is typically done by locating 
sign changes of a real function which has the same modulus as $\zeta$ on the critical line, 
computed using the so-called Riemann--Siegel formula \cite{Siegel1932} and dedicated algorithms, 
such as the Odlyzko--Sch\"onhage algorithm \cite{odlyzko1988fast}, and by counting the zeros 
by means of Turing's method \cite{Turing1953}, see also \cite{Edwards1974}.
It can be used, however, to identify regions in the critical strip which do not contain zeros and 
regions which might contain zeros, and, combined with a counting argument discussed below, 
to prove that the zeros in a given range lie on the critical line.
To see how this can be achieved note that a point $\bar s \in \mathbb{C}$ is a zero of $\zeta(s)$ if and only if
$$
\hat\zeta_n(\bar s) = \bar s R_n(\bar s),
$$
which implies, by inequality (\ref{eq:boundR}), that $\bar s$ is not a zero if
\begin{equation}\label{eq:test}
|\hat\zeta_n(\bar s)|> \frac{|\bar s|\, |\bar s+1|}{8 \left(\Re(\bar s)+1\right)}\, n^{-\Re(\bar s)-1}.
\end{equation}
\rev{The computation of $\hat\zeta_n$ and the exclusion test are summarized, in pseudo-code and as a few lines of Python, 
in ``Computing the approximation and the exclusion test''.}

% ---------- Sidebar pseudo-code (added in the revision) -----------
\begin{sidebar}{Computing the approximation and the exclusion test}\label{sidebar:code}
\revon\sdbarinitial{T}he approximation $\hat\zeta_n(s)$ of equation (\ref{eq:zeta_approx_s}), its error bound in inequality (\ref{eq:boundR}), 
and the exclusion test in inequality (\ref{eq:test}) require only the elementary operations below. 
The recursion runs over $k = 1, \ldots, n-1$ and accumulates $G_n(s) = \sum_{k=1}^{n-1} J_k(s)$.

\medskip
\noindent\textbf{Input:} $s \in \mathbb{C}$ with $\Re(s) > -1$, $s \ne 1$; $n \in \mathbb{N}$.\\
\noindent\textbf{Output:} $\hat\zeta_n(s)$, the bound $B_n(s)$, and the verdict of the test.
\begin{enumerate}
\item $G \leftarrow 0$
\item for $k = 1, \ldots, n-1$: \quad $G \leftarrow G + \ds\frac{(k+1)^{1-s} - k^{1-s}}{1-s} - \left(k+\tfrac{1}{2}\right) \frac{k^{-s} - (k+1)^{-s}}{s}$
\item $\hat\zeta_n \leftarrow \ds\frac{s}{s-1} - \frac{1}{2} - s\,G$
\item $B_n \leftarrow \ds\frac{|s|\,|s+1|}{8\,(\Re(s)+1)}\, n^{-\Re(s)-1}$
\item if $|\hat\zeta_n| > B_n$: $s$ is not a zero of $\zeta$; otherwise: undecided (increase $n$ or refine).
\end{enumerate}

\medskip
\noindent In Python, with the arbitrary precision package \texttt{mpmath} \cite{johansson2017mpmath} (\texttt{mp.dps} sets the number of digits):
\begin{quote}
\texttt{from mpmath import mp, mpf, mpc, fabs}\\
\texttt{mp.dps = 30}\\
\texttt{def zeta\_hat(s, n):}\\
\texttt{\ \ \ \ G = mpc(0)}\\
\texttt{\ \ \ \ for k in range(1, n):}\\
\texttt{\ \ \ \ \ \ \ \ G += ((k+1)**(1-s) - k**(1-s))/(1-s) - (k+mpf(1)/2)*(k**(-s) - (k+1)**(-s))/s}\\
\texttt{\ \ \ \ return s/(s-1) - mpf(1)/2 - s*G}\\
\texttt{def bound(s, n):}\\
\texttt{\ \ \ \ return fabs(s)*fabs(s+1)/(8*(s.real+1)) * n**(-s.real-1)}\\
\texttt{s = mpc(0.6, 14.1); n = 1000}\\
\texttt{print(fabs(zeta\_hat(s, n)) > bound(s, n))\ \ \# True: s is not a zero}
\end{quote}
The cost is $O(n)$ operations per point. Rounding errors are not controlled by this snippet; 
for a certified test they are bounded by evaluating the same expressions in ball arithmetic, see the Appendix.
\end{sidebar}
% ---------- end sidebar -----------

The results of this exclusion test for $n = 10$, which provides very inaccurate results, 
$n=100$, and $n=1000$, which provides very accurate estimates for the location of the nontrivial zeros, 
are illustrated in Figure \ref{fig:nontrivial_zeros}, where the location of the zeros of the Riemann zeta function is 
highlighted with $\star$ markers. In particular, for $n=1000$ each of the 29 zeros with $\omega \in [0,100]$ is contained in a 
region contained in a square of side $0.035$, and all points outside these regions are excluded.
Two comments are in order. First, the test in inequality (\ref{eq:test}) is applied on a grid of points: 
to certify that a whole region does not contain zeros, it has to be complemented by a bound on the derivative of $\hat\zeta_n$, 
or by interval arithmetic, which is available in arbitrary precision packages such as \cite{johansson2017mpmath}. 
Second, in analytic number theory the term {\em zero-free region\ \/}denotes regions of the form 
$\Re(s) > 1 - c/\log|\Im(s)|$, for some $c>0$ \cite{Titchmarsh1986}: the regions determined herein are of a different nature, 
since they are bounded and are computed numerically.

The exclusion test can be turned into a fully certified procedure. This requires three additional ingredients. 
First, the rounding errors in the evaluation of $\hat\zeta_n$ must be bounded, for example by means of interval (ball) arithmetic \cite{Johansson2017arb}. 
Second, the test has to be applied to cells, rather than to points, which requires a bound on the derivative of $\hat\zeta_n$ in each cell. 
Third, each region which is not excluded has to be enclosed in a box, symmetric with respect to the critical line, 
which contains exactly one zero. Since the nontrivial zeros are symmetric with respect to the critical line, 
a zero which is the only zero in such a box necessarily lies on the critical line. Remarkably, the number of zeros in each box 
can be certified by an argument which is familiar in systems theory: if the approximation error is smaller than $|\hat\zeta_n|$ 
on the boundary of the box, then $\zeta$ and $\hat\zeta_n$ have the same number of zeros in the box, exactly as, in the robust 
Nyquist criterion, a perturbation which is smaller than the distance of the nominal Nyquist plot from the origin does not 
change the number of encirclements. A possible implementation of these ideas, and of their extension to large values of $\omega$, 
is outlined in ``Appendix: Toward certified computations at large heights''. 
Note, however, that the cost of the approach is dominated by the cost of each evaluation of $\hat\zeta_n$, which requires $O(n)$ operations, 
and that, by inequality (\ref{eq:boundR}), a fixed accuracy on the critical line at height $\omega$ requires $n$ proportional to $\omega^{4/3}$ 
(or, using higher-order Euler--Maclaurin corrections, proportional to $\omega$). The Riemann--Siegel formula requires instead $O(\sqrt{\omega})$ 
operations per evaluation, which explains why record computations, such as \cite{PlattTrudgian2021}, rely on it. 
As a result, the approximation $\hat\zeta_n$ is well suited to certify the location of the zeros at moderate heights, 
whereas computations at large heights require combining the exclusion and counting arguments with the Riemann--Siegel formula.

\begin{figure*}[htb]
\begin{center}
\centerline{\begin{tikzpicture}
\node[anchor=south west,inner sep=0] (img) at (0,0) {\includegraphics[height=0.78\textheight,keepaspectratio]{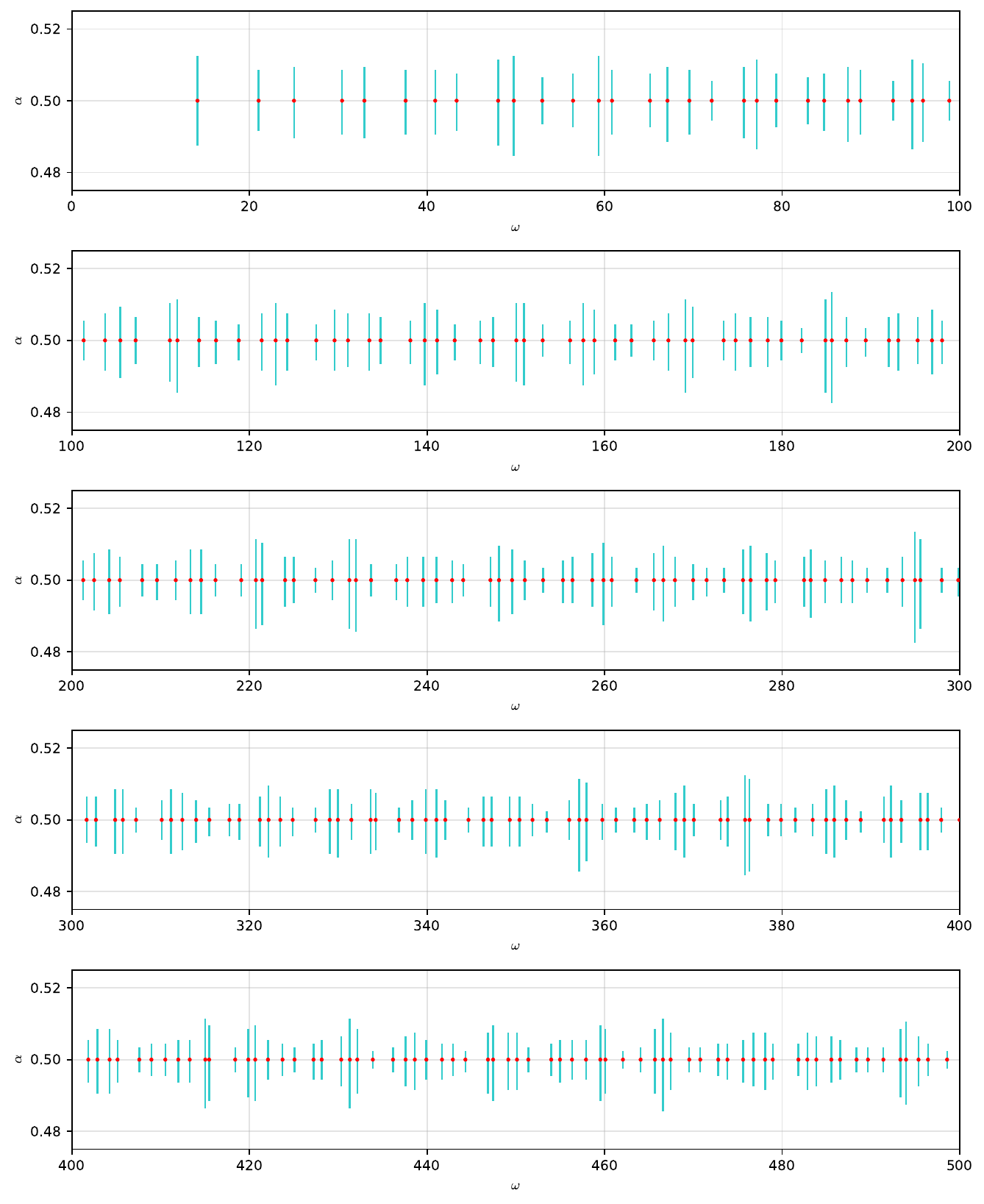}};
\legendbox{($(img.north east)+(-0.6,-0.35)$)} {{\color{lightblue}\rule[-1pt]{10pt}{6pt}}\ not excluded by inequality (\ref{eq:test})\\ {\color{red}$\bullet$}\ zeros of $\zeta$};
\end{tikzpicture}}
\caption{Regions which may contain nontrivial zeros (light blue) for $\alpha \in [0.475, 0.525]$ and $\omega \in [0, 500]$, 
with $n$ selected, in each strip of width $10$, such that the bound in inequality (\ref{eq:boundR}) on the critical line does not exceed $0.02$.  The red markers indicate the locations of the zeros of the Riemann zeta function.
\label{fig:0_500_zeros}}
\end{center}
\end{figure*}

\begin{figure*}[htb]
\begin{center}
\centerline{\begin{tikzpicture}
\node[anchor=south west,inner sep=0] (img) at (0,0) {\includegraphics[height=0.78\textheight,keepaspectratio]{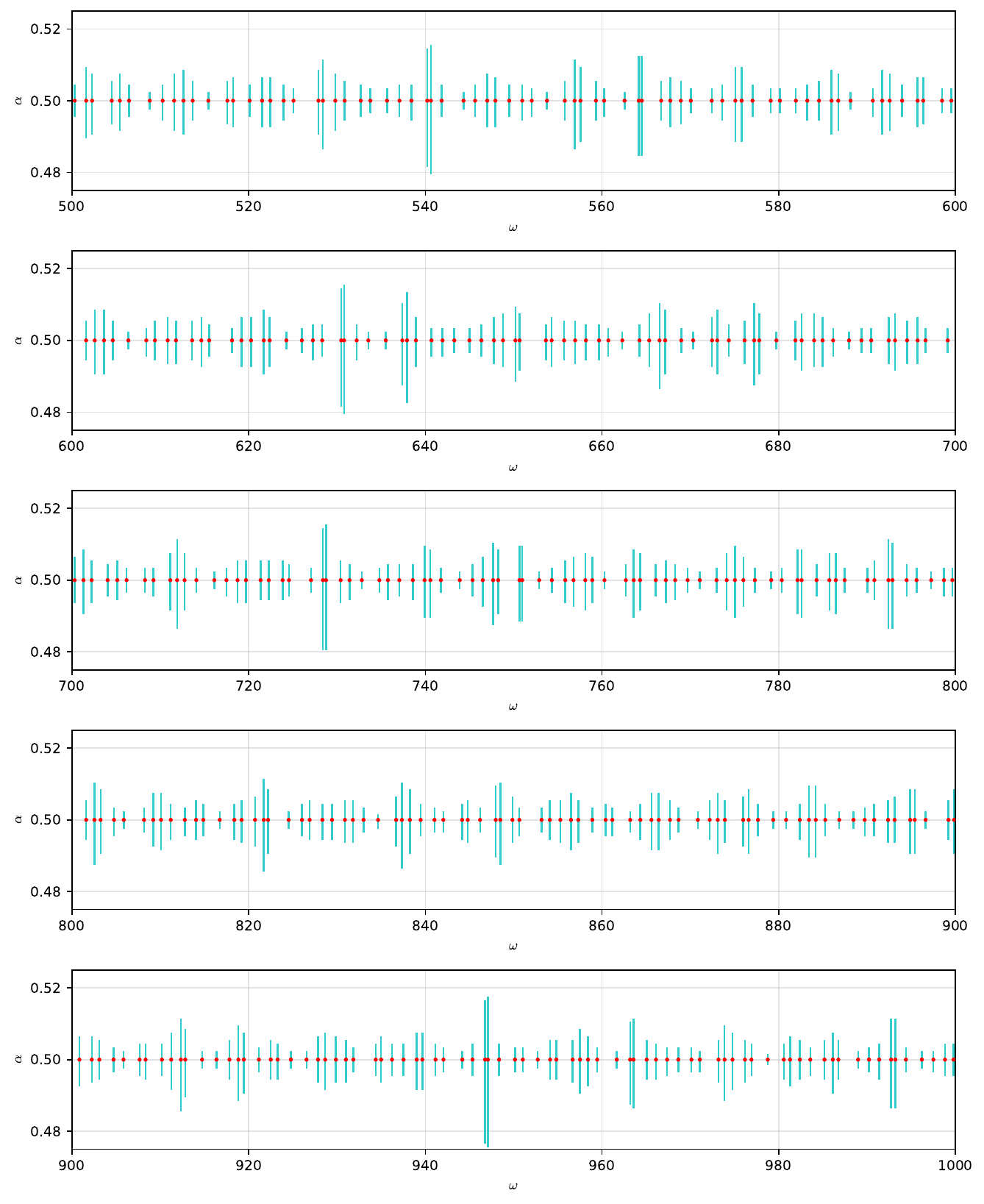}};
\legendbox{($(img.north east)+(-0.6,-0.35)$)} {{\color{lightblue}\rule[-1pt]{10pt}{6pt}}\ not excluded by inequality (\ref{eq:test})\\ {\color{red}$\bullet$}\ zeros of $\zeta$};
\end{tikzpicture}}
\caption{Regions which may contain nontrivial zeros (light blue) for $\alpha \in [0.475, 0.525]$ and $\omega \in [500, 1000]$, 
with $n$ selected, in each strip of width $10$, such that the bound in inequality (\ref{eq:boundR}) on the critical line does not exceed $0.02$.  The red markers indicate the locations of the zeros of the Riemann zeta function.
\label{fig:500_1000_zeros}}
\end{center}
\end{figure*}

An additional illustration of the accuracy of the proposed approximation process is given in 
Figures \ref{fig:0_500_zeros} and \ref{fig:500_1000_zeros},
in which the exclusion test in inequality (\ref{eq:test}) is used to locate all zeros of the 
Riemann zeta function up to $\omega=1000$. Since the bound in inequality (\ref{eq:boundR}) on the critical line grows 
as $\omega^2 n^{-3/2}$, the range of $\omega$ has been divided into strips of width $10$, and in each strip $n$ has been selected 
as the smallest integer such that the bound, evaluated on the critical line at the largest value of $\omega$ in the strip, 
does not exceed $0.02$. This selection, which yields values of $n$ ranging from $57$, for $\omega \in [0,10]$, to about $2.6 \cdot 10^4$, 
for $\omega \in [990,1000]$, makes the size of the regions which are not excluded comparable along the whole range of $\omega$. 
All 649 zeros with $\omega \in [0,1000]$ lie in the regions which are not excluded, each zero lies in its own region, 
and every such region contains a zero.
For comparison, the Riemann Hypothesis has been verified, rigorously, for all zeros with imaginary part up to 
$3 \cdot 10^{12}$ \cite{PlattTrudgian2021}.

\section{The open loop inverse zeta system}

% --------------
\begin{figure*}[htb]
\begin{center}
\begin{tikzpicture}[
    block/.style={draw,rectangle,minimum width=3cm,minimum height=1.5cm,align=center},
    >={Latex}
 ]

\node[block] (B) {$\ds\frac{1}{\zeta(s)}$}; \draw[->, thick] (-5,0) -- (B.west)node[midway, above] {$\ds u(s)= \frac{1}{s}$};
\draw[->, thick] (B.east) -- (5,0) node[midway, above] {$\ds y(s)= {\cal L} M(e^t)$};

\end{tikzpicture}
\caption{The inverse zeta system.}
\label{fig:invZeta}
\end{center}
\end{figure*}
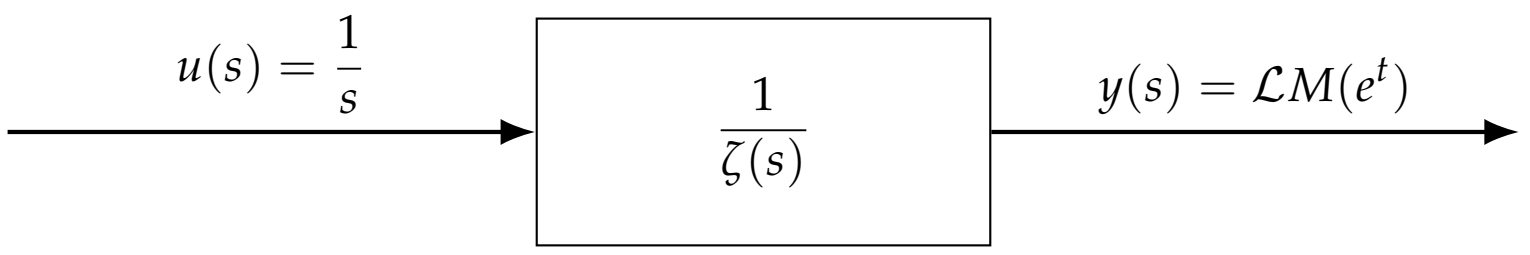
% --------------

To characterize the dominant zero of the Riemann zeta function, consider the function 
$\frac{1}{\zeta(s)}$, that is the multiplicative inverse of the Riemann zeta function,
and regard this as the transfer function of a linear (infinite dimensional) system driven, also in this case, by the function $H$, 
as illustrated in Figure \ref{fig:invZeta}. To begin with, note that, for $\Re(s)>1$,
$$
\frac{1}{\zeta(s)} = \sum_{n \in \mathbb{N}}\frac{\mu(n)}{n^s},
$$
where $\mu$ denotes the M\"obius function,
and that $\frac{1}{\zeta(s)} $ has the inverse Laplace transform
$$
{\cal L}^{-1} \frac{1}{\zeta(s)} = \sum_{n \in \mathbb{N}} \mu(n) \delta(t - \log n),
$$
which (again) allows interpreting the function $\frac{1}{\zeta(s)}$ as the transfer function
of a system composed of the parallel interconnection of countably many delay systems, 
with delay $\tau_n = \log n$ and gain $\mu(n)$, with $n \in \mathbb{N}$.

The output response of the system in Figure \ref{fig:invZeta} with input the Heaviside function
can be computed explicitly to give
$$
y(t) = M(e^t),
$$
that is the Mertens function evaluated at $e^t$.  This function is piecewise constant and such that $|M(e^t)| \le e^t$, hence
it is Laplace transformable, its Laplace transform converges absolutely in the half-plane $\{s \in \mathbb{C}\ |\ \Re(s) >1\}$, and
$$
{\cal L} M(e^t) = \frac{1}{s\,\zeta(s)}, \qquad \Re(s)>1.
$$
The growth of the step response of the inverse zeta system, that is of the Mertens function, 
bounds the location and the multiplicity of the zeros of the Riemann zeta function, as detailed in the following statement.

\begin{proposition}\label{prop:Mertens}
Suppose that there exist $\alpha \in [1/2,1)$, $m \ge 1$, $K>0$, and $t_0 \ge 0$ such that
\begin{equation}\label{eq:bound}
|M(e^t)| \le K\, e^{\alpha t}\, t^{m-1},
\end{equation}
for all $t \ge t_0$. Then the Riemann zeta function does not have zeros in the half-plane $\Re(s)>\alpha$, that is its dominant zero 
is not larger than $\alpha$, and every zero of the Riemann zeta function with real part equal to $\alpha$ has multiplicity not larger than $m$.
\end{proposition}

\begin{IEEEproof}
Inequality (\ref{eq:bound}) implies that the Laplace transform of $M(e^t)$ converges absolutely in the half-plane $\Re(s)>\alpha$, 
hence it defines a function $Y(s)$ which is analytic in this half-plane. Since $Y(s) = \frac{1}{s\zeta(s)}$ for $\Re(s)>1$ and $\zeta$ is 
meromorphic, by uniqueness of the analytic continuation the function $\frac{1}{s\zeta(s)}$ is analytic in the half-plane $\Re(s)>\alpha$, 
that is the Riemann zeta function does not have zeros in this half-plane. Moreover, for all $s = \sigma + i \omega$ with $\sigma>\alpha$,
$$
\begin{array}{ccl}
|Y(s)| & \le & \ds \int_0^{t_0} e^{(1-\sigma)t} dt + K \int_{t_0}^\infty t^{m-1} e^{-(\sigma - \alpha)t} dt \\[4mm]
       & \le & \ds C + \frac{K\,\Gamma(m)}{(\sigma-\alpha)^m},
\end{array}
$$
for some $C>0$ independent of $\sigma \in (\alpha,1]$. Suppose now that $\rho = \alpha + i \omega_0$ is a zero of multiplicity $r$ of the Riemann zeta function. 
Then, as $\sigma$ tends to $\alpha$ from the right, $|Y(\sigma + i\omega_0)| \ge c\, (\sigma-\alpha)^{-r}$, for some $c>0$, which, combined with the 
previous bound, implies $r \le m$.
\end{IEEEproof}

In the language of systems theory, Proposition~\ref{prop:Mertens} is the infinite dimensional counterpart of the familiar fact that a step 
response which grows as $t^{m-1} e^{\alpha t}$ reveals poles with real part not larger than $\alpha$ and with multiplicity not larger than $m$. 
Its significance lies in the fact that it relates the Riemann Hypothesis, and the multiplicity of the nontrivial zeros, to bounds 
on the growth of the Mertens function. 
For example, if for every $\epsilon>0$ there exists $K_\epsilon>0$ such that $|M(x)| \le K_\epsilon x^{1/2+\epsilon}$, for all $x \ge 1$, 
then the Riemann Hypothesis holds: this is one direction of a classical characterization of the Riemann Hypothesis, see \cite[Chapter 14]{Titchmarsh1986}. 
Similarly, a bound of the form $|M(x)| \le K \sqrt{x} (\log x)^{m-1}$ would imply the Riemann Hypothesis and that all nontrivial zeros 
have multiplicity not larger than $m$. In particular, the validity of a conjecture of Gonek, see \cite{ng2004distribution}, which states that
$M(x) = O\!\left(\sqrt{x} (\log\log\log x)^{5/4}\right)$, would imply, for every $\epsilon>0$, inequality (\ref{eq:bound}) with $\alpha=1/2$ and $m = 1 + \epsilon$, 
hence the Riemann Hypothesis and the simplicity of the nontrivial zeros of the Riemann zeta function.
Note, finally, that inequality (\ref{eq:bound}) with $\alpha=1/2$ and $m=1$, that is $M(x) = O(\sqrt{x})$, is believed to be false, 
see \cite{OdlyzkoTeRiele1985,KotnikTeRiele2006,ng2004distribution,hurst2016computations}: 
growth conditions of the form considered in Proposition~\ref{prop:Mertens} are sufficient, but not necessary, 
for the location of the zeros of the Riemann zeta function.

% ---------- Sidebars Mobious and Mertens -----------

\begin{sidebar}{The M\"obius function}\label{def:Mobius}
\sdbarinitial{T}he M\"obius function $\mu$, see \cite{Apostol1976,HardyWright2008,Mobius1832},
is defined, for all $n \in \mathbb{N}$, as 
\begin{itemize}
\item $\mu(1) = 1$;
\item $\mu(n) = 0$ if $n$ has a squared prime factor;
\item $\mu(n) = (-1)^k$ if $n$ is a product of $k$ distinct primes.
\end{itemize}
The M\"obius function is a multiplicative function which plays a key role in multiplicative number theory through 
the M\"obius inversion formula, 
which allows one to recover an arithmetic function from its summatory transform. 
It is closely connected to the distribution of prime numbers and the Riemann zeta function via the identity
$$
\frac{1}{\zeta(s)}=\sum_{n=1}^{\infty}\frac{\mu(n)}{n^s}, \qquad \Re(s)>1.
$$
\end{sidebar}

\begin{sidebar}{The Mertens function}\label{def:Mertens}
\sdbarinitial{T}he Mertens function $M$, see \cite{Apostol1976,HardyWright2008,Mertens1897},
is defined, for all $n \in \mathbb{N}$, as
$$
M(n) = \sum_{k=1}^n \mu(k).
$$
The Mertens function can be extended to the reals by setting
\begin{equation}\label{eq:Mreal}
M(x)=M(\lfloor x\rfloor ).
\end{equation}
The Mertens function is an important function in number theory: 
it measures the balance between positive and negative values of the Möbius function and is closely connected to the distribution of prime numbers. 
The growth of $M$ has been extensively studied because of its relationship to the zeros of the Riemann zeta function. 
A famous conjecture, the Mertens Conjecture, claimed that $|M(n)|<\sqrt{n}$ for all $n \in \mathbb{N}$, 
but it was disproved in 1985 by Andrew Odlyzko and Herman te Riele \cite{OdlyzkoTeRiele1985}, 
who showed that $\limsup_{x\to\infty} M(x)/\sqrt{x} > 1.06$. It is believed that $M(x)/\sqrt{x}$ is unbounded \cite{ng2004distribution}.
The Mertens function satisfies the identity
$$
\sum_{k=1}^{n} M\!\left( \left\lfloor \frac{n}{k} \right\rfloor \right) = 1,
$$
for all $n \in \mathbb{N}$, see \cite{Apostol1976}.
\end{sidebar}

% -------

\section{The feedback zeta system}

\begin{figure*}[htb]
\begin{center}
\begin{tikzpicture}[
    block/.style={draw,rectangle,minimum width=2.5cm,minimum height=1.2cm,align=center},
    sum/.style={draw,circle,minimum size=0.8cm},
    >={Latex}
]

% Nodes
\node[sum] (S) at (-5,0) {};
\node[block] (K) at (-2,0) {$K$};
\node[block] (B) at (2,0) {$\zeta(s)$};

% Input r
\draw[->, thick] (-8,0) -- (S.west)
    node[midway, above] {$\!\!\!\!\!\! \ds r(s)=\frac{1}{s}$};

% Forward path
\draw[->, thick] (S.east) -- (K.west);
\draw[->, thick] (K.east) -- (B.west)
    node[midway, above] {$u(s)$};

% Output
\draw[->, thick] (B.east) -- (8,0)
    node[midway, above]
    {$\displaystyle y(s) = \zeta_{cl}(s) \ \frac{1}{s} $};

% Feedback path
\draw[->, thick]
    (B.east) -- ++(1,0)
    |- (4,-3)
    -| (S.south);

% Signs at summing junction
\node at (-5.75,0.25) {$+$};
\node at (-5.50,-0.75) {$-$};

\end{tikzpicture}
\caption{Closed-loop zeta system with gain $K$ and negative feedback.}
\label{fig:ZetaClosedLoop}
\end{center}
\end{figure*}
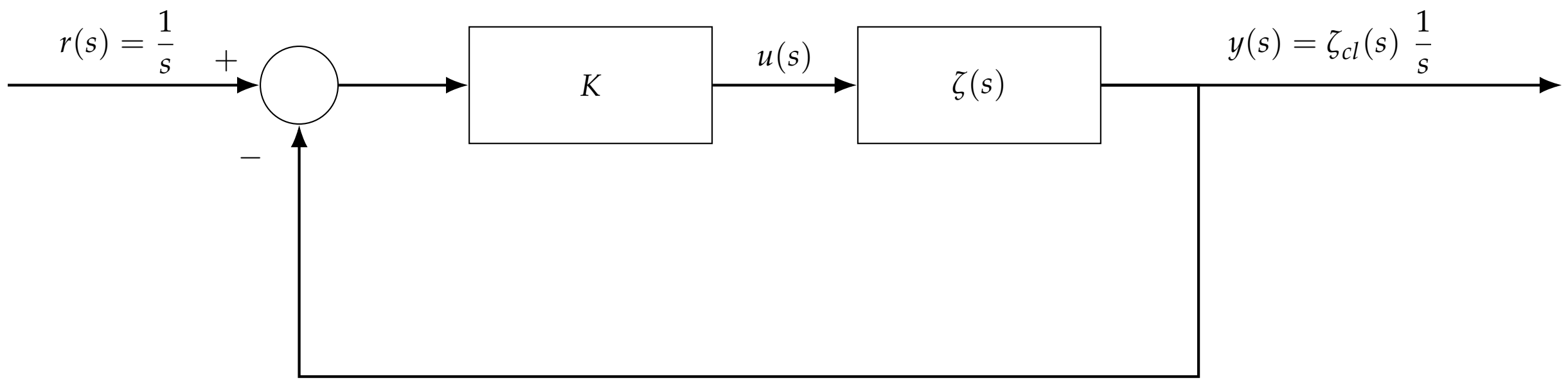

From a control theory perspective, it is natural to consider the zeta system in a feedback loop, 
as illustrated in Figure \ref{fig:ZetaClosedLoop}, with a static gain $K>0$, and to study its behavior for large $K$.

\begin{figure}[htb]
\begin{center}
\centerline{\includegraphics[width=20.0pc]{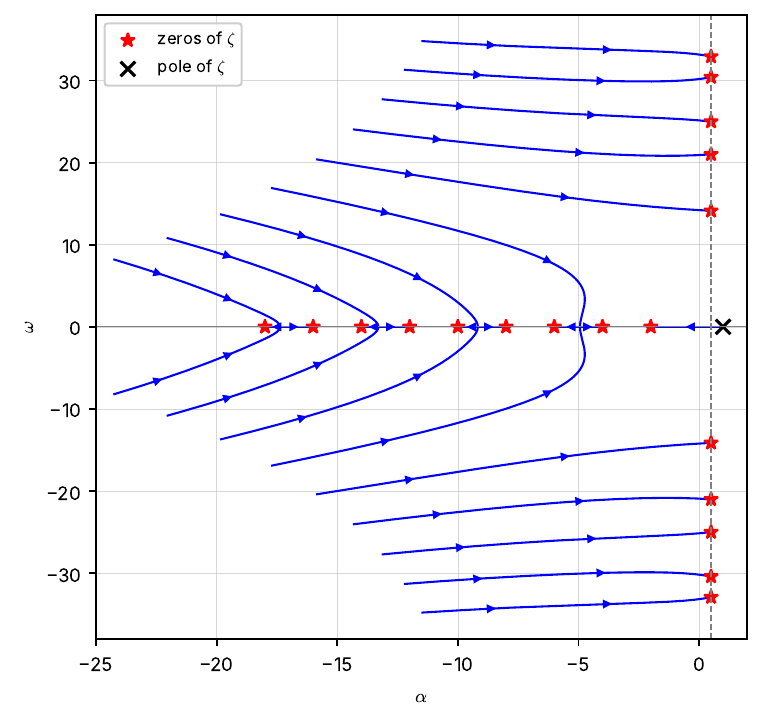}}
\caption{\rev{The closed-loop poles, that is the solutions of $\zeta(s) = -1/K$, for $K$ between $10^{-9}$ and $10^4$, 
which converge to the first five nontrivial zeros of the Riemann zeta function, to their complex conjugates, and to the first nine trivial zeros. 
The arrows indicate the direction of increasing $K$. The dashed line is the critical line.}
\label{fig:closed_loop_poles}}
\end{center}
\end{figure}

To this end, note that $\zeta(s)$ tends to $1$ as $\Re(s)$ tends to $+\infty$, that is the zeta system has a direct feedthrough 
(the impulse at $t = \log 1 = 0$), hence the closed-loop system is well-posed for all $K>0$. Note now that
$$
u(s) = K (r(s) - y(s)), \qquad y(s) = \zeta(s) u(s),
$$
hence
$$
y(s) = \frac{K \zeta(s)}{1+ K \zeta(s)} \ \frac{1}{s} = \zeta_{cl}(s) \ \frac{1}{s}, \qquad
u(s) = \frac{K}{1+ K \zeta(s)} \ \frac{1}{s}.
$$
Observe that the poles of the closed-loop system are such that
$$
\zeta(s) = -\frac{1}{K},
$$
hence as $K$ tends to $+\infty$ the poles of the closed-loop transfer function $\zeta_{cl}(s)$ tend to the zeros of the Riemann zeta function. 
\rev{This statement refers to the finite zeros of the Riemann zeta function and is made precise as follows. 
Let $\rho$ be a zero of $\zeta$ of multiplicity $m$ and let $D$ be a disk centered at $\rho$ which contains no other zero of $\zeta$. 
Since $1 + K\zeta(s)$ converges to $K \zeta(s)$ uniformly on $D$, relative to $|K\zeta(s)|$, as $K$ tends to $+\infty$, 
Rouch\'e's theorem implies that, for all $K$ sufficiently large, the function $1+K\zeta(s)$ has exactly $m$ zeros in $D$, 
and these converge to $\rho$ as $K$ tends to $+\infty$ (this is Hurwitz's theorem, see \cite{Ahlfors1979}). Conversely, if $s_K$ is a closed-loop pole and $s_K$ converges to a finite point $s^\star$ 
as $K$ tends to $+\infty$, then $\zeta(s_K) = -1/K$ converges to zero, hence, by continuity, $\zeta(s^\star) = 0$. 
As a result, the finite limit points of the closed-loop poles are exactly the zeros of the Riemann zeta function; 
closed-loop poles may also tend to infinity, which is the case, for instance, as $K$ tends to zero. 
Figure \ref{fig:closed_loop_poles} displays the closed-loop poles, computed numerically for $K$ between $10^{-9}$ and $10^4$, 
which converge to the first five nontrivial zeros, to their complex conjugates, and to the trivial zeros between $-18$ and $-2$. 
As $K$ increases the poles approach the nontrivial zeros along almost horizontal paths, 
and for $K=100$ they are within $0.02$ of the critical line. 
The trivial zeros are reached along the real axis. Near $s=1$ the Riemann zeta function behaves as $\frac{1}{s-1}$, 
hence one closed-loop pole leaves the pole of the zeta function, at $s=1$, and moves along the real axis 
to the trivial zero at $s=-2$, crossing the origin for $K=2$, since $\zeta(0) = -\frac{1}{2}$. 
The remaining trivial zeros are reached in pairs: on the real axis the Riemann zeta function is negative in the intervals $(-4k-2,-4k)$ 
and positive in the intervals $(-4k,-4k+2)$, with $k \in \mathbb{N}$, hence, for $K$ sufficiently large, the equation $\zeta(s) = -\frac{1}{K}$ 
has two real solutions in each interval $(-4k-2,-4k)$, one close to each of its end points. 
As $K$ increases, a complex conjugate pair of closed-loop poles reaches the real axis at the minimum of the Riemann zeta function in the interval, 
for $K$ approximately equal to $251$, $127$, $10.7$, and $0.27$, for $k=1,2,3,4$, respectively, 
and then splits into two real poles, which converge to the end points of the interval, exactly as at a break-in point of a classical root locus. 
As $K$ tends to zero, all closed-loop poles, with the exception of the one which converges to $s=1$, 
move to the left and tend to infinity, along curves on which, for $\alpha$ negative and sufficiently large in modulus, 
$|\omega|$ grows as $\frac{\pi |\alpha|}{2 \log(|\alpha|/(2\pi))}$, so that their direction tends to that of the negative real axis. 
This follows from the functional equation (S2) and from Stirling's formula, since $\zeta(1-s)$ tends to $1$ 
as $\Re(s)$ tends to $-\infty$. 
The figure is the analogue, for the zeta system, of the root locus of a finite dimensional system, 
in which the closed-loop poles migrate from the open-loop poles to the open-loop zeros as the gain increases. 
Note that, for a finite dimensional system, the branches of the root locus which do not start at finite poles 
come from infinity along asymptotes; for the zeta system the point at infinity is an essential singularity, 
hence no such simple description of the behavior at infinity is available.}

Moreover, for all $s$ such that $\zeta(s) \neq 0$, 
$$
\lim_{K \to \infty} u(s) = \frac{1}{s\, \zeta(s)},
$$ 
that is, as is well known for high-gain feedback, the closed-loop system asymptotically inverts the zeta system, 
and the control signal $u$ tends to the step response of the inverse zeta system.
This property can be established, rigorously, in the time domain. 
Since the impulse response of the zeta system is a sum of impulses at the times $\log k$, with $k \in \mathbb{N}$, 
and $\log k + \log m = \log(km)$, the signals $u$ and $y$ are constant in the intervals $[\log n, \log(n+1))$, with $n \in \mathbb{N}$.
Setting $u_n = u(\log n)$ and $y_n = y(\log n)$, and noting that $y(\log(n/k)) = y_{\lfloor n/k \rfloor}$, yields
$$
y_n = \sum_{k=1}^{n} u_{\lfloor n/k \rfloor}, \qquad u_n = K (1 - y_n),
$$
for all $n \in \mathbb{N}$. Eliminating $y_n$ yields the recursion
\begin{equation}\label{eq:yn2}
(1+K)\, u_n = K \left( 1 - \sum_{k=2}^{n} u_{\lfloor n/k \rfloor} \right),
\end{equation}
which determines $u_n$, and then $y_n$, explicitly for any $K>0$, since $\lfloor n/k \rfloor < n$ for all $k \ge 2$. 
For each $n$, taking the limit as $K$ tends to $+\infty$ in equation (\ref{eq:yn2}) yields
$$
\sum_{k=1}^{n} \lim_{K \to \infty} u_{\lfloor n/k \rfloor} = 1, \qquad \lim_{K \to \infty} y_n = 1.
$$
The former relation determines the limits of the $u_n$'s uniquely and coincides with the identity satisfied by the Mertens function, 
see ``The Mertens function''. As a result,
$$
\lim_{K \to \infty} u(t) = M(e^t), \qquad \lim_{K \to \infty} y(t) = 1, 
$$
for all $t \in \mathbb{R}_{\ge 0}$.
In summary, in the high-gain limit the output tracks the reference exactly: the closed-loop poles converge to the zeros of the 
Riemann zeta function, but their contribution to the output vanishes. The zeros are revealed, instead, by the control signal, 
whose growth is characterized by Proposition~\ref{prop:Mertens}.

% ------------ end sidebars 

\section{CONCLUSIONS}

Some properties of the Riemann zeta function - a fundamental object in 
analytic number theory - have been studied from a systems and control perspective. 
The point of departure has been the interpretation of the Riemann zeta function, and its multiplicative inverse, 
as the transfer functions of linear, infinite dimensional, delay systems.
It has been shown that the Mertens function appears naturally as the step response of the inverse of the so-called 
{\em zeta system}, and as the limit of the control signal of the zeta system in a high-gain feedback loop. 
This has allowed exploiting a systems and control theory perspective, and very simple tools, namely the Laplace transform and its region of convergence,
to derive properties of the Riemann zeta function;
to derive algorithms for the computation of the function, which coincide with first-order Euler--Maclaurin summation, with explicit error bounds; 
to compute regions of the critical strip which do not contain zeros;
and to relate the growth of the Mertens function to the location and the multiplicity of the zeros of the Riemann zeta function.
\rev{None of these facts is new to number theory: the value lies in the elementary and unified derivation, 
and in the logical structure of the certified algorithm of the Appendix, a first implementation of which 
has certified all zeros with imaginary part up to $10^8$, see the Appendix.}

\rev{The dictionary works in both directions, and it is worth asking what systems theory may take back from number theory. 
First, the Mellin transform and the theory of Dirichlet series provide a mature toolkit for 
transfer functions with countably many poles, such as those of delay and other distributed parameter systems \cite{Curtain2009}: 
the abscissae of convergence of Dirichlet series are exactly the growth exponents of step responses, and the explicit, 
quantitative versions of Perron's formula give explicit bounds on the responses in terms of the pole-zero distribution. 
Second, the exclusion and counting arguments of the Appendix, that is a certified test for the absence of zeros in a cell 
and the Rouch\'e--Nyquist count of the zeros in a box, apply verbatim to the characteristic function of any 
delay system, and provide a certified alternative to the numerical computation of its spectrum, 
with explicit regions of the complex plane which are proved to contain no poles. 
Third, the identities of multiplicative number theory, of which the one satisfied by the Mertens function in the high-gain limit 
is an example, describe the input-output behavior of the zeta system at the sampling instants $\log n$, 
and suggest that the parallel interconnections of delay systems whose delays are logarithms of integers 
form a class of systems whose analysis can borrow, systematically, from arithmetic.}

\section{Appendix: Toward certified computations at large heights}

This appendix outlines how the exclusion test in inequality (\ref{eq:test}) could be developed into an algorithm which, 
given $0<T_1<T_2$, certifies that all zeros of the Riemann zeta function with imaginary part in $[T_1,T_2]$ lie on the critical line 
and are simple, and determines their number. Such an algorithm would verify the Riemann Hypothesis in the window $[T_1,T_2]$. 
The algorithm can be implemented using parallel computing, since different windows, and different cells within a window, can be processed independently, 
and it is therefore suited to an implementation on a supercomputer. In addition, its output is a {\em certificate}, namely a list of 
cells and of computed enclosures, which can be stored and checked independently.

\subsection{Certified evaluation}

Let $F$ be a computable approximation of the Riemann zeta function such that 
\begin{equation}\label{eq:app_B}
|\zeta(s) - F(s)| \le B(s),
\end{equation}
for all $s$ in the region of interest, where $B$ is an explicit function. Two choices are available. 
For moderate heights one can select $F = \hat\zeta_n$ and $B$ equal to the right-hand side of inequality (\ref{eq:boundR}). 
For large heights one can select the Riemann--Siegel approximation, for which explicit error bounds are available on the 
critical line \cite{Gabcke1979} and in the critical strip \cite{AriasdeReyna2011}. 
In both cases $F$ is evaluated in ball arithmetic \cite{Johansson2017arb}: each computed quantity is returned as a midpoint and a radius, 
and the exact value is guaranteed to lie in the corresponding disk, rounding errors included. Since the evaluation of $\hat\zeta_n$ 
requires the summation of $n$ terms, the working precision has to grow (mildly) with $n$.

\subsection{Certified exclusion of cells}

The following statement allows excluding whole cells, rather than points.

\begin{proposition}\label{prop:cell}
Let $C \subset \mathbb{C}$ be a closed convex set, let $c \in C$, and let $r = \max_{s \in C} |s-c|$. 
Suppose that $F$ is analytic in a neighborhood of $C$ and that
\begin{equation}\label{eq:cell_test}
|F(c)| > \max_{s \in C} B(s) + r \max_{s \in C} |F^\prime(s)|.
\end{equation}
Then the Riemann zeta function does not have zeros in $C$, and $|F(s)| > B(s)$, for all $s \in C$.
\end{proposition}

\begin{IEEEproof}
Since $C$ is convex, for all $s \in C$ the segment joining $c$ and $s$ is contained in $C$, hence 
$|F(s)| \ge |F(c)| - r \max_{s \in C} |F^\prime(s)| > \max_{s \in C} B(s) \ge B(s)$. 
As a result, by inequality (\ref{eq:app_B}), $|\zeta(s)| \ge |F(s)| - B(s) > 0$, for all $s \in C$.
\end{IEEEproof}

For $F = \hat\zeta_n$ the quantities in inequality (\ref{eq:cell_test}) are explicit. In fact, the derivatives of $\hat\zeta_n$ are
$$
\begin{array}{ccl}
\hat\zeta_n^{(j)}(s) & = & \ds \sum_{k=2}^{n-1} (-\log k)^j\, k^{-s} + \frac{(-\log n)^j\, n^{-s}}{2} \\[5mm]
                     &   & \ds + \, n^{1-s} \sum_{i=0}^{j} \binom{j}{i} \frac{(-\log n)^{j-i} (-1)^i\, i!}{(s-1)^{i+1}},
\end{array}
$$
hence, if $\Re(s) \ge \sigma_0$ for all $s \in C$ and $d$ denotes the distance between $C$ and the point $s=1$, 
$|\hat\zeta_n^{(j)}(s)| \le M_j$, for all $s \in C$, with
$$
\begin{array}{ccl}
M_j & = & \ds \sum_{k=2}^{n-1} k^{-\sigma_0} (\log k)^j + \frac{n^{-\sigma_0} (\log n)^j}{2} \\[5mm]
    &   & \ds + \, n^{1-\sigma_0} \sum_{i=0}^{j} \binom{j}{i} \frac{(\log n)^{j-i}\, i!}{d^{i+1}}.
\end{array}
$$
The bound $M_1$ on $\max_{s \in C} |\hat\zeta_n^\prime(s)|$ is, however, very pessimistic near the critical line, 
since it grows as $\sqrt{n} \log n$. A much sharper bound is obtained by expanding $\hat\zeta_n^\prime$ at the center of the cell, which yields
$$
\max_{s \in C} |\hat\zeta_n^\prime(s)| \le \sum_{j=1}^{K-1} \frac{|\hat\zeta_n^{(j)}(c)|\, r^{j-1}}{(j-1)!} + \frac{M_K\, r^{K-1}}{(K-1)!},
$$
for any integer $K \ge 2$. Note that the derivatives at $c$ are computed in the same pass over $k$ as $\hat\zeta_n(c)$, 
hence the cost of each cell test is $O(Kn)$, and that the pessimistic term is multiplied by $r^{K-1}$.
Finally, $\max_{s \in C} B(s)$ is bounded by evaluating the right-hand side of inequality (\ref{eq:boundR}) with $\Re(s)$ replaced by $\sigma_0$ 
and $|s|$ replaced by $\max_{s \in C} |s|$. Cells for which inequality (\ref{eq:cell_test}) fails are subdivided, for example by bisection, 
until the test succeeds or the cells become smaller than a prescribed size.

\subsection{Certified isolation of zeros on the critical line}

The following statement exploits the symmetry of the zeros and Rouch\'e's theorem.

\begin{proposition}\label{prop:box}
Let $Q = [\frac{1}{2}-\delta, \frac{1}{2}+\delta] \times [a,b]$, with $0 < \delta < \frac{1}{2}$ and $0<a<b$, 
be a box in the complex plane, and suppose that $F$ is analytic in a neighborhood of $Q$. 
Suppose that the boundary $\partial Q$ of $Q$ is covered by cells which satisfy inequality (\ref{eq:cell_test}). 
Then the Riemann zeta function and $F$ have the same number of zeros in $Q$, counted with their multiplicity. 
If this number is equal to one, then the Riemann zeta function has exactly one zero in $Q$, this zero is simple, and it lies on the critical line.
\end{proposition}

\begin{IEEEproof}
By Proposition~\ref{prop:cell} and inequality (\ref{eq:app_B}), $|\zeta(s) - F(s)| \le B(s) < |F(s)|$ for all $s \in \partial Q$. 
Since $\zeta$ is analytic in a neighborhood of $Q$ (note that $1 \notin Q$), by Rouch\'e's theorem $\zeta$ and $F$ have the same number of 
zeros in $Q$. Suppose now that this number is equal to one, and let $\rho = \beta + i \gamma$ be the zero of $\zeta$ in $Q$, which is 
therefore simple. By the functional equation (\ref{eq:functional_eq}) and the relation $\zeta(\bar s) = \overline{\zeta(s)}$, 
the point $1-\beta + i\gamma$ is also a zero of $\zeta$, and it belongs to $Q$, since $Q$ is symmetric with respect to the critical line. 
Hence $1 - \beta = \beta$, that is $\beta = \frac{1}{2}$.
\end{IEEEproof}

The number of zeros of $F$ in $Q$ is computed, rigorously, by means of the argument principle. To this end, $\partial Q$ is divided 
into segments such that, on each segment, the ball enclosure of $F$ lies in an open half-plane whose boundary contains the origin. 
On each such segment the variation of the argument of $F$ is uniquely determined, and belongs to $(-\pi,\pi)$, and the sum of these 
variations, divided by $2\pi$, yields the number of zeros of $F$ in $Q$. The cells used in Proposition~\ref{prop:box} play the role 
of the uncertainty disks in a robust Nyquist test: the {\em nominal\ \/}function $F$ determines the number of encirclements, and the 
bound $B$ guarantees that the {\em true\ \/}function $\zeta$ has the same number. Note that, for the Riemann--Siegel approximation, 
the number of terms of the main sum changes when $\sqrt{\Im(s)/(2\pi)}$ crosses an integer, hence the boxes have to be selected 
so that $F$ is analytic in a neighborhood of each of them.

\subsection{The algorithm}

Combining the above statements yields the following procedure, to be executed for each window $[T_1,T_2]$.

\medskip

\begin{enumerate}
\item Locate, non-rigorously, candidate zeros $\frac{1}{2} + i \gamma_j$, with $\gamma_j \in [T_1,T_2]$, for example by detecting sign changes 
      of Hardy's function $Z(t) = e^{i\theta(t)} \zeta(\frac{1}{2}+it)$, which is real for real $t$ \cite{Edwards1974}.
\item Enclose each candidate in a box $Q_j$, symmetric with respect to the critical line, with size of the order of the bound $B$ divided by 
      $|F^\prime(\frac{1}{2}+i\gamma_j)|$, multiplied by a safety factor.
\item Cover the rectangle $[\frac{1}{2},1] \times [T_1,T_2]$, with the exception of the interiors of the boxes $Q_j$, by cells which 
      satisfy inequality (\ref{eq:cell_test}), refining adaptively. The values $T_1$ and $T_2$ are selected so that the corresponding 
      horizontal segments are not contained in any box.
\item Compute the number of zeros of $F$ in each box $Q_j$ by the argument principle, and verify that it is equal to one.
\end{enumerate}

\medskip

If all steps succeed then, by Propositions~\ref{prop:cell} and \ref{prop:box}, by the symmetry of the zeros with respect to the critical line, 
and since $\zeta$ does not have zeros with $\Re(s) \ge 1$, all zeros with imaginary part in $[T_1,T_2]$ lie on the critical line, are simple, 
and their number is equal to the number of boxes. If a cell outside the boxes cannot be excluded, then either the accuracy of the computation 
(the value of $n$, or the working precision) is insufficient, or a zero off the critical line has been found, which would disprove the 
Riemann Hypothesis: the two cases can be distinguished by increasing the accuracy. If the number of zeros in a box is larger than one, 
the box contains two close zeros, or a multiple zero: in the former case the box is split into smaller boxes and the computation is repeated 
with higher accuracy. Note that, differently from the classical approach based on Turing's method \cite{Turing1953}, which requires evaluations 
on the critical line only, the procedure requires evaluations in a two-dimensional region: this is more expensive, but it is conceptually simple, 
and it provides, in addition, explicit regions which do not contain zeros.

\subsection{Computational cost}

The number of zeros with imaginary part in a window of width $W$ at height $T$ is approximately $\frac{W}{2\pi} \log\frac{T}{2\pi}$, 
and the total cost is the product of this number, of the number $E$ of cell tests per zero, and of the cost of each test. 

To assess $E$, the covering procedure has been simulated, in double precision (hence without rigorous control of the rounding errors), 
for windows of width $W=10$ at heights $T=10^2$, $T=10^3$, and $T=10^4$, with $F = \hat\zeta_n$, $n$ selected so that $B \le 0.02$ on the critical line, 
boxes with half-size equal to $2B/|F^\prime(\frac{1}{2}+i\gamma_j)|$, initial cells of side $1/4$, and bisection. 
In all cases every cell outside the boxes has been excluded. With $K=4$ the number of cell tests per zero has been $E = 180$, $E=308$, and $E=556$, 
respectively, whereas with $K=6$ it has been $E=69$, $E=98$, and $E=135$, respectively. 
Replacing the factor $2$ in the size of the boxes with $1.5$ or $3$ has changed these values by about $10$ percent. 
These counts do not include the (few) additional evaluations which may be required by the argument principle. 
Hence, at moderate heights and with a suitable order $K$, $E$ is of the order of $10^2$ and grows slowly with $T$.
Whether this remains true at much larger heights has not been verified: the bounds on the derivatives, and hence the size of the cells, 
depend on $n$ and on the approximation used. The estimates below are therefore expressed in terms of $E$.
\newmat{The implementation described at the end of this Appendix, which uses the Riemann--Siegel approximation, 
has measured $E$ between $47$ and $58$ at all heights between $10^2$ and $10^9$.}

The cost of each evaluation depends dramatically on the selected approximation. 
For $F = \hat\zeta_n$ and a fixed bound $B$ on the critical line, inequality (\ref{eq:boundR}) requires $n$ proportional to $T^{4/3}$, 
and higher-order Euler--Maclaurin corrections reduce this to $n$ proportional to $T$. 
For the Riemann--Siegel approximation the main sum has approximately $\sqrt{T/(2\pi)}$ terms, and the cost can be further reduced, 
when many evaluations are required, by the Odlyzko--Sch\"onhage algorithm \cite{odlyzko1988fast}.

To fix ideas, consider first a window of width $W=10^3$ at height $T=10^6$, which contains about $1.9 \cdot 10^3$ zeros. 
With $F = \hat\zeta_n$ and $B \le 0.02$ on the critical line, $n \approx 2.6 \cdot 10^8$, and the computation requires about $5 \cdot 10^{11} E$ terms, 
that is between $5 \cdot 10^{13}$ and $5 \cdot 10^{14}$ terms for $E$ between $10^2$ and $10^3$, which is within reach of a cluster. 
With the Riemann--Siegel approximation the main sum has only about $400$ terms. 
Consider now a window of width $W = 10^4$ at height $T = 10^{20}$, which contains about $7 \cdot 10^4$ zeros. 
With $F = \hat\zeta_n$ each evaluation would require at least $10^{19}$ terms, which is not feasible. 
With the Riemann--Siegel approximation each evaluation requires about $4 \cdot 10^9$ terms, and the computation requires about $3 \cdot 10^{14} E$ terms, 
that is between $3 \cdot 10^{16}$ and $3 \cdot 10^{17}$ terms for $E$ between $10^2$ and $10^3$, each evaluated in multiple precision: 
this is a substantial, but feasible, computation on a modern supercomputer, and its cost can be reduced considerably by exploiting the Odlyzko--Sch\"onhage algorithm.

In summary, the systems theoretic ideas discussed in this paper, namely the exclusion test, the robust (Rouch\'e-Nyquist) counting 
argument, and the symmetry of the zeros, provide the logical structure of a certified algorithm, whereas the efficiency of the algorithm at 
large heights relies on the classical Riemann--Siegel formula and on fast multi-evaluation algorithms. 
Note finally that a verification of the Riemann Hypothesis beyond the current record \cite{PlattTrudgian2021} requires processing all 
windows up to the selected height, and not only windows at large heights: for such a task the classical approach, which only requires 
evaluations on the critical line, is more efficient, and the procedure described above is best regarded as an independent, 
and complementary, certification tool.

\newmat{\subsection{A first implementation}
At the time of writing, the procedure has been implemented, in C with the Arb library \cite{Johansson2017arb}, and run on the 
Shaheen III supercomputer at KAUST \cite{Astolfi2026report}. The implementation uses the Riemann--Siegel approximation 
as evaluated by Arb, with its rigorous error bound, and obtains the Taylor coefficients needed for the cell test from a discrete 
Cauchy integral, that is from point evaluations on a small circle around the center of the cell, with the aliasing error 
bounded by means of a rigorous bound on $|\zeta|$ over a disk; every quantity is a ball, so that the output is a certificate 
in the sense described above, which is re-verified by a separate program. 
With this implementation all zeros of the Riemann zeta function with imaginary part in $[2, 10^8]$, that is $248{,}008{,}025$ zeros, 
have been certified to lie on the critical line and to be simple, in agreement with the known count, with the certificates 
(one per window of width $100$) independently re-checked on a sample; windows at heights $10^9$, $10^{10}$, and $10^{12}$ 
have been certified as well. The cost has been about $2 \cdot 10^5$ core-hours for the range $[2,10^8]$, 
with $47$ to $58$ cell tests per zero and between $10$ and $70$ milliseconds per test, and the certification 
of the range up to $10^9$ is in progress. The code and the certificates will be released under an open licence.}

\section{Author Information}

\begin{IEEEbiography}{{A}lessandro Astolfi}{\,}(alessandro.astolfi@kaust.edu.sa)
(IFAC Fellow, FIEEE, Academia Europaea, FIET, ITATEC, ACA Fellow, EUCA Fellow) 
received the Laurea in Electronic Engineering from the University of Rome La Sapienza, Italy, 
in 1991; the M.Sc. degree in Information Theory and the Ph.D. degree with Medal of 
Honor with a thesis on discontinuous stabilization of nonholonomic systems from ETH-Zurich, Switzerland, in 1995; 
and the Ph.D. degree for his work on nonlinear robust control from the University of Rome “La Sapienza” in 1996.

From 1992 to 1996 he was a Research Associate at ETH-Zurich. 
From 1996 to 2026 he was with the Electrical and Electronic Engineering Department, Imperial College London. 
In particular, he was Professor of Nonlinear Control Theory; College Consul for the Faculty of Engineering and Business School (2022-2025); 
and Head of the Control and Power Group (2010-2022). 
He retains a part-time appointment as Professor of Nonlinear Control Theory within the Dyson School of Design Engineering at Imperial College London. 
From 1998 to 2003 he was an Associate Professor with the Department of Electronics and Information, Politecnico of Milano. 
From 2005 to 2025 he was a Professor with the Dipartimento di Ingegneria Civile e Ingegneria Informatica, University of Rome Tor Vergata, Rome. 
Since February 2026, he has been a Professor of Applied Mathematics and Computational Science at KAUST, Saudi Arabia.

His research interests include mathematical control theory and control applications, with special emphasis for the 
problems of discontinuous stabilization, robust and adaptive control, observer design, optimal control, game theory, and model reduction.

Dr. Astolfi was the recipient of the IEEE CSS A. Ruberti Young Researcher Prize (2007); 
the IEEE RAS Googol Best New Application Paper Award (2009); 
the IEEE CSS George S. Axelby Outstanding Paper Award (2012); 
the Automatica Best Paper Award (2017); 
and the IEEE Transactions on Control Systems Technology Outstanding Paper Award (2023). 
He is a “Distinguished Member” of the IEEE CSS and a recipient of the IFAC Outstanding Service Award (2017). 
He is the recipient of the Institute of Measurement and Control Sir Harold Hartley Medal for 
“outstanding contributions to the technology of measurement and control”.

He was Editor-in-Chief of the IEEE Transactions on Automatic Control (2018–2025) 
and Chair of the IEEE CSS Conference Editorial Board (2010–2017). 
He is Vice Chair of the IFAC Technical Board (2020–2026); 
and a Member of the IEEE Thesaurus Editorial Board (2025–) and of the IEEE PRAC (2026–).
\end{IEEEbiography}

% --------------------------

% ---------------
% ---------------

% ---------------
% ---------------

\balance

% --------------------------

\endarticle

\end{document}